\documentclass[american,aps,prx,superscriptaddress,longbibliography,floatfix]{revtex4-2}

\usepackage{inputenc}
\usepackage{bbold}
\usepackage{color}
\usepackage{babel}
\usepackage{mathrsfs}
\usepackage{physics}
\usepackage{array} 
\usepackage{mathtools}
\usepackage{bbding}
\usepackage{pifont}
\usepackage{textcomp}
\usepackage{wasysym}
\usepackage{amsthm}
\usepackage{nicefrac}
\usepackage{wasysym}
\usepackage{amsmath}
\usepackage{bbm}
\usepackage{amssymb}
\usepackage{graphicx}
\usepackage{enumitem}
\usepackage{xcolor}

\usepackage[colorlinks=true,linkcolor=blue,citecolor=red,plainpages=false,pdfpagelabels]{hyperref}
\usepackage{booktabs}   
\usepackage{tabularx}

\newtheorem{theorem}{Theorem}
\newtheorem*{theorem*}{Theorem}

\newtheorem{proposition}{Proposition}
\newtheorem*{proposition*}{Proposition}
\newtheorem{corollary}{Corollary}
\newtheorem{lemma}{Lemma}
\newtheorem*{example*}{Example}

\newtheorem*{remark*}{Remark}
\newtheorem{definition}{Definition}
\newtheorem*{lemma*}{Lemma}

\def\tr{\operatorname{tr}}

\def\St{\operatorname{St}}
\def\Ch{\operatorname{Ch}}
\def\SCh{\operatorname{SCh}}
\def\Pos{\operatorname{Pos}}
\def\id{\operatorname{id}}

\def\n{\mathcal{N}}
\newcommand{\mc}[1]{\mathcal{#1}}

\begin{document}

\title{Thermodynamic work capacity of quantum information processing}

\author{Himanshu Badhani}\email{himanshubadhani@gmail.com}
\affiliation{q4i, Centre for Quantum Science and Technology, International Institute of Information Technology, Hyderabad, Gachibowli, Telangana 500032, India}
\affiliation{Center for Security, Theory and Algorithmic Research, International Institute of Information Technology, Hyderabad, Gachibowli, Telangana 500032, India}

\author{Dhanuja G.S.}
\affiliation{q4i, Centre for Quantum Science and Technology, International Institute of Information Technology, Hyderabad, Gachibowli, Telangana 500032, India}
\affiliation{Center for Computational Natural Sciences and Bioinformatics, International Institute of Information Technology, Hyderabad, Gachibowli, Telangana 500032, India}

\author{Siddhartha Das}
\email{das.seed@iiit.ac.in}
\affiliation{q4i, Centre for Quantum Science and Technology, International Institute of Information Technology, Hyderabad, Gachibowli, Telangana 500032, India}
\affiliation{Center for Security, Theory and Algorithmic Research, International Institute of Information Technology, Hyderabad, Gachibowli, Telangana 500032, India}

\begin{abstract}
We introduce the resource-theoretic free energy of a quantum channel as the maximal work extractable from the channel as its output equilibrates to a thermal state and its reference system remains locally intact. It is proportional to the relative entropy between the given channel and the absolutely thermal channel. It attains a clear operational meaning as twice the asymptotic rates of athermality distillation and formation under Gibbs preserving superchannels, which map one absolutely thermal channel to another for a given bath, thereby revealing the asymptotic reversibility of the resource theory of athermality for quantum channels. Consequently, we establish that the optimal extractable work in converting one channel to another through the asymptotic athermality distillation and formation tasks equals the difference in their free energies. We call this optimal work the thermodynamic work capacity of channel conversion. Quantum information processing and computing fundamentally concern the manipulation and transformation of quantum channels, which encompass quantum states, their transformations, and measurements. A quantitative characterization of the optimal thermodynamic work gain or expenditure in quantum information processing constitutes a key step toward formulating thermodynamics of quantum processes.
\end{abstract}

\maketitle
\twocolumngrid
Understanding the thermodynamic aspects of quantum systems and processes is crucial for studying natural phenomena and developing energy-efficient quantum technologies~\cite{Sag12,Auf22,Pot24,BD26}. Thermodynamic effects impact the informational and computational capabilities of quantum processors, and the ongoing drive toward miniaturization increases the relevance of quantum thermodynamics~\cite{GMN+15,AH19,SS21,SDC21,BGD25}. Information theory provides the bridging conceptual notions between thermodynamics and quantum theory, along with tools and techniques to build the frameworks for quantum thermodynamics and analyze the energetics of quantum systems and processes~\cite{Ben03,Gou24,DS25,HWS+25}.

Quantum information processing and computing primarily rely on physical transformations of quantum channels~\cite{CDP09,Das19,DBWH21,SPSD25}, which are completely positive, trace-preserving linear maps. Furthermore, the study of thermodynamical and informational aspects of quantum channels seem to readily provide a unifying framework to understand these aspects also for quantum states and measurements~\cite{DGP24,BD26}, as they are special examples of quantum channels. A quantum state is a preparation channel that acts on $1$ and outputs the state. A quantum measurement is a quantum to classical channel that acts on quantum states to yield classical outputs with some probability distribution.

The free energy and work extraction are basic concepts in thermodynamics. We want to find appropriate and meaningful definitions of these concepts for quantum channels that would shed light on the thermodynamics of quantum information processing. It will help us perceive the thermodynamic utility of quantum channels and the energetics of their implementations in quantum devices. The extension of these concepts to quantum channels need not be straightforward~\cite{NG15,FBB21,DGP24,BGC+25} as the complications seem to get manifested in the notions involving channel discrimination tasks~\cite{Wat05,BCR11,DW19b,FFRS20,SPSD25}. 

We identify the meaningful definitions of free energy and work extraction from quantum channels, and develop the resource theory of athermality of quantum channels (see also \cite{BGD25,BD26}). The resource-theoretic free energy of a quantum channel is defined as the maximum work that can be extracted when the channel output is thermalized in the presence of a bath, while the reference to the channel remains locally intact. We find that the resource-theoretic free energy of the channel is equal to the channel relative entropy with respect to the absolutely thermal channel (up to a multiplicative factor of the inverse temperature). An absolutely thermal channel always outputs a fixed thermal state with inverse temperature associated to the bath~\cite{DS25}. This motivates us to develop a resource theory of athermality of channels with respect to a bath at a fixed, arbitrary temperature where the free object is an absolutely thermal channel and free operations are Gibbs preserving superchannels (GPSCs). For a given thermal bath, a Gibbs preserving superchannel maps one absolutely thermal channel to another at the same inverse temperature.

The relative entropy of athermality of a quantum channel is the relative entropy between the given channel and the absolutely thermal channel. It is an athermality monotone quantifying how athermal the channel is. It finds an exact operational meaning as twice the asymptotic athermality distillation and formation rates of the channel, thereby exhibiting that the resource theory of athermality of quantum channels under the allowed action of Gibbs preserving superchannels is asymptotically reversible. As a consequence, the optimal work extractable during the asymptotic conversion of a quantum channel to another under Gibbs preserving superchannels, is equal to the difference in their free energy. We call this optimal extractable work as the \textit{thermodynamic work capacity} of the conversion of quantum channels.

\textit{Prior works}.--- Consider a quantum system $A$ with a Hamiltonian $\widehat{H}_A$ that is weakly interacting with the bath, such that the total energy of the system-bath can be well approximated as the sum of the individual energies. Then, the equilibrium state of the system $A$ is given by the thermal (Gibbs) state $\gamma^\beta_A:=\exp(-\beta \widehat{H}_A)/Z^{\beta}_A$~\cite{Len78}, where $\beta$ is the inverse temperature of the bath and $Z^{\beta}_A:=\tr \exp(-\beta \widehat{H}_A)$ is the partition function. The nonequilibrium free energy of a state $\rho_A$ is defined as $F^{\beta}_{\rm T}(\rho):=E(\rho)-\beta^{-1}S(\rho)$ (cf.~\cite{BHO+13,Gou24,DC19}), where  $S(\rho):=-\tr(\rho\ln\rho)$ is the von Neumann entropy and $E(\rho)=\tr(\rho \widehat{H}_A)$ is the internal energy of the system. We refer to $F^{\beta}_{\rm T}(\rho)$ as the thermodynamic free energy of the state $\rho$. For the thermal state, the equilibrium free energy is $F_{\rm T}^{\beta}(\gamma^\beta_A)=-\beta^{-1}\ln Z^{\beta}_A$. Incidentally, this quantity is related to the quantum relative entropy of the state with the thermal state, $F^{\beta}_{\rm T}(\rho)=\beta^{-1}D(\rho\Vert\gamma^{\beta})+F_{\rm T}^{\beta}(\gamma^\beta)=\beta^{-1} D(\rho\Vert\widehat{\gamma}^\beta)$~\cite{BHO+13,BHN+15}, where $D(\rho\Vert\sigma):=\tr (\rho\ln\rho-\rho\ln\sigma)$, and $\widehat{\gamma}^\beta:=\exp(-\beta \widehat{H}_A)$ is the thermal operator for given $\widehat{H}_A$ and $\beta$. 

The amount of work $W$ that can be extracted as we evolve from one equilibrium state $\gamma^\beta_1$ to another $\gamma^\beta_2$ is bounded by the change in the thermodynamic free energy, $W\le F_{\rm T}^{\beta}(\gamma^\beta_2)-F_{\rm T}^{\beta}(\gamma^\beta_1)$, where the inequality saturates if the process is reversible. The maximum amount of work that can be extracted from a nonequilibrium state $\rho$ as the system equilibrates to the thermal state $\gamma^\beta$, is given by $W^\beta_{\rm ext}(\rho\to\gamma^\beta)=\beta^{-1}D(\rho\Vert\gamma^\beta)$~\cite{EV11} when asymptotically many copies of $\rho$ are available. $F^\beta(\rho):=\beta^{-1}D(\rho\Vert\gamma^\beta)$ is called the resource-theoretic free energy of a quantum state $\rho$~\cite{BHO+13,Gou24}. The maximum extractable work through the conversion of a state $\rho$ to a state $\sigma$ under the action of Gibbs preserving channels is given by $W^\beta_{\rm ext}(\rho\to\sigma)=F^\beta_{\rm T}(\rho)-F^\beta_{\rm T}(\sigma)=F^\beta(\rho)-F^\beta(\sigma)$ when asymptotically many copies of $\rho$ and $\sigma$ are available. This is a direct consequence of the fact that the resource theory of athermality of quantum states is asymptotically reversible~\cite{CG19,Gou22,Lam25} under the action of Gibbs preserving channels. The Gibbs preserving channels map a thermal state to a thermal state with the same inverse temperature; they don't generate athermality in quantum states.

The utility of a class of athermal channels as thermal engines by the distillation of work from them has been discussed in a different resource-theoretic framework in \cite{NG15,FBB21} by considering unitary operations that commute with the thermal state as free operations. In these works, the possible use of reference to the channel input and output is ignored. On the contrary, it is desirable to consider unitary operations as athermal resources as they are purity-preserving and exhibit genuine quantum behavior~\cite{BGC+25,DGP24} unlike the absolutely thermal channels~\cite{DS25}. Furthermore, the task of quantum channel discrimination informs us that the reference to the channel is often crucial and can provide an advantage, which otherwise would not be feasible. From (quantum) thermodynamics, we know that even correlation carries information apart from the possible storage of information on local states. The extractable work from a quantum correlated state $\rho_{AB}$ is more than the extractable work from the product state $\rho_A\otimes\rho_B$. Under a global entropy preserving process, when the bath and systems are weakly interacting, the maximal extractable work from a state $\rho_{AB}$ is given~\cite{BHN+15,BPF+15,PHH+15,BRL+19}
\begin{align}
    W^\beta_{\rm ext}(\rho_{AB})&=\beta^{-1}I(A;B)_{\rho}+F^\beta(\rho_A)+F^\beta(\rho_B)\\
    &=F^\beta_{\rm T}(\rho_{AB})-F_{\rm T}^{\beta}(\gamma^\beta_A\otimes\gamma^\beta_B)\\
    &=\beta^{-1}D(\rho_{AB}\|\gamma^\beta_A\otimes\gamma^\beta_B)=F^\beta(\rho_{AB}),
\end{align}
where we have assumed that $\widehat{H}_{AB}=\mathbb{1}_A\otimes\widehat{H}_B+\widehat{H}_A\otimes\mathbb{1}_B$ and $I(A;B)_{\rho}:=D(\rho_{AB}\|\rho_A\otimes\rho_B)$ is the quantum mutual information of the state $\rho_{AB}$, and for a state $\rho_{AB}$, we let $\rho_A:=\tr_B(\rho_{AB})$.

\textit{Extractable work from a quantum channel}.--- We would like to consider the optimal work extractable from a quantum channel $\n_{A'\to A}$ through a thermodynamic quantity that depends on how well the channel can preserve the correlation between its input $A'$ and the reference and while also accounting for the (resource-theoretic) free energy of the channel output $A$. To capture this idea, we define the extractable work from a quantum channel as the maximum possible work extractable during the partial thermalization process $\id_R\otimes\mc{N}_{A'\to A}(\psi_{RA'})\to\psi_R\otimes\gamma^\beta_A$ under global operations when systems weakly interact with the bath,  and optimize it over all possible input states $\psi\in\St(RA')$ for arbitrary dimension $R$,
\begin{align}
    &W^{\beta}_{\rm ext}[\n]
    \nonumber \\
    &:=\sup_{\psi\in\St(RA')}\left[F^\beta_{\rm T}(\id\otimes\mc{N}(\psi_{RA'}))-F^\beta_{\rm T}(\psi_R\otimes\gamma^\beta_A)\right]\\
    &=\sup_{\psi\in\St(RA')}\left[F^\beta(\id\otimes\mc{N}(\psi_{RA'}))-F^\beta(\psi_R)\right]\label{eq:max-w-ch}\\
&=\sup_{\psi\in\St(RA')}\left[\beta^{-1}I(R;A)_{\id\otimes\n(\psi)}+F^\beta(\n(\psi_{A'}))\right],
\end{align}
where we assumed that $\widehat{H}_{RA}=\mathbb{1}_R\otimes\widehat{H}_A+\widehat{H}_R\otimes\mathbb{1}_A$. We advocate that the Eq.~\eqref{eq:max-w-ch} can be interpreted as extractable work from the channel $\n$ in similar way as $S[\n]:=\inf_{\psi\in\St(RA')}[S(\id\otimes\n(\psi_{RA'}))-S(\psi_R)]$~\cite[Eq.~(1.1)]{DJKR06} finds meaningful interpretation as the entropy of the channel $\n$~\cite{GW21,BGC+25,DS25,BGD25,BD26}.

Any transformation $\id_R\otimes\mc{N}_{A'\to A}(\psi_{RA'})\to \psi_R\otimes\gamma^\beta_A$ for all input states $\psi_{RA'}$ is equivalent to the thermalization of the channel $\mc{N}$ itself, i.e., $\mc{N}\to\mc{T}^\beta$, where $\mc{T}^\beta_{A'\to A}(\cdot):=\tr(\cdot)\gamma^\beta_A$ is an absolutely thermal channel~\cite{DS25,BGD25} that always outputs state in thermal equilibrium with a bath at inverse temperature $\beta$. The channel transformation $\mc{N}\to \mc{T}^\beta$ is equivalent to thermalization of the channel output while the reference system being locally intact. We can interpret $W^{\beta}_{\rm ext}[\n]$ as the extractable work when we transform the channel $\n$ to $\mc{T}^\beta$. It follows that $W^{\beta}_{\rm ext}[\n]\geq 0$ for an arbitrary quantum channel $\n$ and the inequality saturates if and only if $\n$ is absolutely thermal, $\n=\mc{T}^\beta$.

\textit{Dynamical free energy}.--- We find that the extractable work for an arbitrary quantum channel $\n_{A'\to A}$ is proportional to the channel relative entropy between the channel and the absolutely thermal channel,
\begin{align}
   W^{\beta}_{\rm ext}[\n] &=\beta^{-1}D[\n\|\mc{T}^\beta]\\
   &=\beta^{-1}(D[\n\Vert\widehat{\mc{T}}^{\beta}]-D[\mc{T}^\beta\Vert \widehat{\mc{T}}^{\beta}]), 
\end{align}
where $\widehat{\mc{T}}^\beta_{A'\to A}(\cdot)=\tr(\cdot)\widehat{\gamma}^\beta_A$ and
\begin{align}
 D[\n\|\mc{M}]=\sup_{\psi_{RA'}}D(\mc{N}(\psi_{RA'})\Vert\mc{M}(\psi_{RA'})),   
\end{align} 
is the quantum relative entropy between completely positive maps $\n_{A'\to A}, \mc{M}_{A'\to A}$ and it suffices to optimize over pure states $\psi_{RA'}$ with $R\simeq A'$. We arrive at the aforementioned identities by observing that $I(A;B)_{\rho}+D(\rho_B\Vert\gamma^\beta)=D(\rho_{AB}\|\rho_A\otimes\gamma^\beta_B)$ and the fact that $D[\mc{T}^\beta\Vert \widehat{\mc{T}}^{\beta}]=D(\gamma^\beta_A\Vert\widehat{\gamma}^\beta_A)=-\ln Z^\beta_A$. The function $F^\beta[\n]:=\beta^{-1}D[\n\|\mc{T}^\beta]$ (and $F^\beta_{\rm T}[\n]:=\beta^{-1}D[\n\|\widehat{\mc{T}}^\beta]$) exhibit the following properties~\cite{BGD25}: (a) monotonically nonincreasing under the action of Gibbs preserving superchannels $\Theta^\beta$, $\Theta^\beta(\mc{T}^\beta_{A'\to A})=\mc{T}^\beta_{B'\to B}$, (b) reduction to the free energy of a state $\omega_A$ for a replacer (or preparation) channel $\mc{R}^{\omega}$, where $\mc{R}^{\omega}_{A'\to A}(\cdot):= \tr(\cdot)\omega_A$, (c) minimum if and only if the channel is absolutely thermal, (d) continuity, (e) additive under tensor-product of quantum channels $\mc{N}_{A'\to A}\otimes\mc{M}_{B'\to B}$ if channel outputs are noninteracting $\widehat{H}_{AB}=\widehat{H}_A\otimes\mathbbm{1}_B+\mathbbm{1}_A\otimes\widehat{H}_B$, $F^\beta[\mc{N}\otimes\mc{M}]=F^\beta[\mc{N}]+F^\beta[\mc{M}]$, (f) convexity. These are desirable properties for a free energy function of a quantum channel, analogous to quantum states (see Proposition~\ref{prop:free-eenrgy-properties} in Appendix~\ref{sec:prop} for proof).

Based on the exhibition of the aforementioned desired properties, we identify $F^\beta[\n]=\beta^{-1}D[\n\|\mc{T}^\beta]$ and $F^\beta_{\rm T}[\n]=\beta^{-1}D[\n\|\widehat{\mc{T}}^\beta]$ as the resource-theoretic and thermodynamic free energy of a quantum channel $\n_{A'\to A}$, respectively, so that 
\begin{equation}\label{eq:w-ext}
W^{\beta}_{\rm ext}[\n]=F^\beta[\n]=F^\beta_{\rm T}[\n]-F^\beta_{\rm T}[\mc{T}^\beta].
\end{equation}
This exemplifies $W^\beta_{\rm ext}[\n]=F^\beta[\n]$ as a faithful monotone of the athermality of a quantum channel $\n$, i.e., a valid monotone of how athermal the channel is and being zero if and only if the channel is absolutely thermal. The utility of an athermal quantum channel, a channel that is not absolutely thermal, for thermodynamic work extraction motivates us to formulate the resource theory of quantum channels.

\textit{Dynamical resource theory of athermality}.--- In real world, often the quantum systems eventually get in equilibrium with the surrounding which acts as a thermal bath. The equilibrium state of the system $A$ that is weakly interacting with the bath at some arbitrary inverse temperature $\beta$, is given by the thermal state $\gamma^\beta_A=\exp(-\beta\widehat{H}_A)/Z^\beta_A$. This suggests that the absolutely thermal channel $\mc{T}^\beta$ can be considered natural or a free object~\cite{DS25}. For a thermal bath at inverse temperature $\beta$, the absolutely thermal channel $\mc{T}^{\beta}$ is unique and any preparation channel $\mc{T}^{\beta'}$ is athermal if $\beta'\neq \beta$. Gibbs preserving superchannels form free operations in the resource theory of athermality as they don't increase athermality of a quantum resource channel and preserves the thermalization property of the free object for a given thermal bath. Based on this observation, we formulate the resource theory of athermality for finite-dimensional square quantum channels under Gibbs preserving superchannels as free operations~\footnote{A quantum channel is said to be square if its input and output dimensions are equal.}. 

\textit{Golden units of athermality}. For a thermal bath at inverse temperature $\beta$, we denote a quantum resource channel $\n_{A'\to A}$, where $|A'|=|A|<+\infty$, as $(\n,\mc{T}^\beta)$. We observe that the relative entropy of athermality $D[\n\|\mc{T}^\beta]$ of a quantum channel $\n$ is maximum if and only if it is a unitary channel. All unitary channels $\mc{U}_{A'\to A}$ are equally resourceful (see Appendix~\ref{subsec:goldenunit}), and interconversions between these unitaries exist under Gibbs preserving superchannels, which can be depicted as $(\mc{U},\mc{T}^\beta)\sim (\id,\mc{T}^\beta)$. It can also be shown that the relative entropy of athermality $D[\mc{U}\|\mc{T}^\beta]$ of a unitary channel $\mc{U}_{A'\to A}$ is minimum if and only if the output Hamiltonian is fully degenerate, $\widehat{H}_A\propto \mathbbm{1}_A$ (see Appendix~\ref{subsec:goldenunit}, Proposition~\ref{prop:min-free-en}). As the output Hamiltonian of a quantum channel can be arbitrary, we advocate a golden unit of athermality of quantum channels to be a unitary channel with a fully degenerate output Hamiltonian. Without loss of generality, we choose the identity channel $(\id_{m},\mc{R}^\pi)$ to be a golden unit for the resource theory of athermality of quantum channels acting on an $m$-dimensional system, where $\id_m$ is the identity channel on an $m$-dimensional system and $\mc{T}^\beta$ reduces to the uniformly mixing channel $\mc{R}^\pi(\cdot)=\tr(\cdot)\pi$, where $\pi$ is the maximally mixed state, for $\widehat{H}\propto \mathbbm{1}$.

\textit{Athermality distillation and formation}. For the transformation of an arbitrary resource channel $(\mc{N}_{A'\to A},\mc{T}^{\beta}_{A'\to A})$ to another arbitrary resource channel $(\mc{M}_{B'\to B},\mc{T}^{\beta}_{B'\to B})$, the conversion distance under the action of Gibbs preserving superchannels (GPSCs) is defined as 
\begin{align}
d_{\rm GP}&((\mc{N},\mc{T}^{\beta})\to (\mc{M},\mc{T}^{\beta}))\nonumber \\
=&\inf_{\Theta}\left\{P[\mc{M},\Theta(\mc{N})]: \Theta(\mc{T}^\beta_{A'\to A})=\mc{T}^\beta_{B'\to B}\right\},    
\end{align}
where $P[\mc{N},\mc{M}]:=\sup_{\psi_{RA'}}P(\mc{N}(\psi),\mc{M}(\psi))$ is the purified distance between quantum channels $\mc{N}_{A'\to A},\mc{M}_{A'\to A}$ and optimization can be taken over pure states $\psi_{RA'}$, and $|A'|=|A|$, $|B'|=|B|$. For an error $\varepsilon\in[0,1]$ and a resource quantum channel $(\n,\mc{T}^\beta)$, the one-shot athermality distillation $\mathrm{Dist}^\varepsilon (\n,\mathcal{T}^\beta)$ and formation $\mathrm{Cost}^\varepsilon (\n,\mathcal{T}^\beta)$ under GPSCs are respectively defined as
\begin{align}
      &  \mathrm{Dist}^\varepsilon (\n,\mathcal{T}^\beta) \nonumber\\&:=\sup_{m}\left\{\ln m:~d_{\mathrm{GP}}((\n,\mathcal{T}^\beta)\rightarrow(\id_m,\mathcal{R}^\pi))\le\varepsilon\right\},\\
     &   \mathrm{Cost}^\varepsilon (\n,\mathcal{T}^\beta)\nonumber\\&:=\inf_{m}\left\{\ln m:~    d_{\mathrm{GP}}((\id_m,\mathcal{R}^\pi)\rightarrow(\n,\mathcal{T}^\beta))\le\varepsilon\right\}.
    \end{align}
That is, up to an allowed error $\varepsilon\in[0,1]$, $\mathrm{Dist}^\varepsilon (\n,\mathcal{T}^\beta)$ is the maximum number of basic golden resource units $(\id_{\rm e},\mc{R}^\pi)$ distillable from $(\mc{N},\mc{T}^\beta)$ and $\mathrm{Cost}^\varepsilon (\n,\mathcal{T}^\beta)$ is the minimum number of basic golden resource units $(\id_{\rm e},\mc{R}^\pi)$ required to expend to form $(\mc{N},\mc{T}^\beta)$. For asymptotically many parallel uses of the resource channel $(\mc{N},\mc{T}^\beta)$ and $\varepsilon\in(0,1)$, the asymptotic distillation and formation rates are the same, 
\begin{align}\label{eq:distillation-formation}
   &\limsup_{n\to\infty}\frac{1}{n}\mathrm{Dist}^\varepsilon (\n^{\otimes n},{\mathcal{T}^\beta}^{\otimes n})\nonumber\\
   =&\liminf_{n\to\infty}\frac{1}{n}\mathrm{Cost}^\varepsilon (\n^{\otimes n},{\mathcal{T}^\beta}^{\otimes n})\\
   =&\frac{1}{2}D[\n\|\mc{T}^\beta],
\end{align}
implying that the resource theory of athermality of quantum channels is reversible. Here, we consider only square quantum channels for the interconversion problem; the extension of this result to arbitrary quantum channels is discussed in \cite{BD26}.

The proof follows by first showing (see Appendix \ref{subsec:distillation}, Theorem~\ref{thm:dist_cost}) that $\mathrm{Distill}^\varepsilon (\n,\mathcal{T}^\beta)=\frac{1}{2}D^{\varepsilon^2}_{H}[\n\|\mc{T}^\beta]$ and $\mathrm{Cost}^\varepsilon (\n,\mathcal{T}^\beta)=\frac{1}{2}D^\varepsilon_{\infty}[\n\Vert\mc{T}^\beta]$~\cite{BGD25} (also see \cite{BD26}), where $D^\varepsilon_H[\n\Vert\mc{M}]$ and $D^\varepsilon_\infty[\n\|\mc{M}]$ are smoothed hypothesis-testing and max relative entropies between channels $\n,\mc{M}$~\cite{CMW16,LKDW18}, respectively. To show these relations, we used the fact that the uniformly mixing channel $\mc{R}^\pi_{A'\to A}$ can be expressed as the uniform mixture of Weyl unitary channels $\mc{W}^i_{A'\to A}$, $\mc{R}^\pi(\cdot)=\frac{1}{m^2}\sum_{i=0}^{m^2-1}\mc{W}^i(\cdot)$ for $|A'|=|A|=m$, where $\mc{W}^i(\cdot):=W^i(\cdot)(W^i)^\dag$ for the set (group) $\{W^i\}_{i=0}^{m^2-1}$ of Weyl unitaries, and $\norm{\id-\mc{R}^\pi}_{\diamond}=1-\frac{1}{m^2}$ ($\norm{\mc{M}}_{\diamond}:=\sup_{\psi_{RA'}}\norm{\id_R\otimes\mc{M}(\psi_{RA'})}_1$ is the diamond norm for a Hermiticity-preserving map $\mc{M}_{A'\to A}$). We then use the asymptotic equipartition property (see Appendix~\ref{subsec:distillation}, Theorem~\ref{thm:rev} for smoothed hypothesis-testing and max-relative relative entropies $D_{\infty}$ between a quantum channel and a replacer channel~\cite{CMW16,FGR25}, $\forall~\varepsilon\in(0,1)$
\begin{align}
    \lim_{n\to \infty}\frac{1}{n}D^\varepsilon_H[\n^{\otimes n}\|{\mc{R}^\omega}^{\otimes n}]&= \lim_{n\to \infty}\frac{1}{n}D^\varepsilon_\infty[\n^{\otimes n}\|{\mc{R}^\omega}^{\otimes n}]\nonumber \\
    &=D[\n\|\mc{R}^\omega],
\end{align}
to arrive at Eq.~\eqref{eq:distillation-formation}.

It is known that the identity channel $\id_{A'\to A}$ allows for the perfect quantum communication, equivalent to the distribution of a maximally entangled state $\Phi_{RA}=\sum_{i,j=0}^{m-1}\ket{ii}\bra{jj}_{RA}$ of Schmidt rank $m$, where $|R|=|A'|=|A|=m$. Operationally, establishing the identity channel $\id_{A'\to A}$ between two parties allows for the perfect distribution of $m\times m$-dimensional quantum states. A golden unit $(\id_{m},\mc{R}^{\pi})$ of dynamical athermality  corresponds to a golden unit $(\Phi_{m},\pi_R\otimes\pi_A)\sim (\op{m}\otimes\op{m},\pi\otimes\pi)$ of static athermality of $m\times m$-dimensional quantum system, where $\op{m}$ is a pure energy eigenstate. This shows that $(\id_{m},\mc{R}^{\pi})$ is equivalent to the double of $(\op{m},\pi_{m})$, where $\pi_{m}=\frac{1}{m}\mathbbm{1}$, in terms of athermality~\cite{BGD25}.

\textit{Thermodynamic work capacity of channel conversion}.--- We define the one-shot net extractable work $W^{\varepsilon_1,\varepsilon_2}_{\beta}[\n\to\mc{M}]$ for thermodynamic channel transformation from an arbitrary square quantum channel $\mc{N}_{A'\to A}$ to an arbitrary square quantum channel $\mc{M}_{B'\to B}$ as the surplus (gain) in extractable work during the one-shot athermality distillation from $(\mc{N},\mc{T}^\beta)$ up to an error $\varepsilon_1\in[0,1]$ and the one-shot athermality formation of $(\mc{M},\mc{T}^\beta)$ up to an error $\varepsilon_2\in[0,1]$, both under GPSCs,
\begin{align}
   W^{\varepsilon_1,\varepsilon_2}_{\beta,{\rm ext}}[\n\to\mc{M}]= \frac{2}{\beta}\left[\mathrm{Dist}^{\varepsilon_1} (\n,\mathcal{T}^\beta)-\mathrm{Cost}^{\varepsilon_2} (\mc{M},\mathcal{T}^\beta)\right].
\end{align}
The thermodynamic work capacity $\Delta W^\beta_{\rm ext}[\mc{N}\to \mc{M}]$ of the channel conversion $\mc{N}\to\mc{M}$ is defined as the difference between the asymptotic work distillation rate from $(\mc{N},\mc{T}^\beta)$ and the asymptotic work cost rate of $(\mc{M},\mc{T}^\beta)$, which is equal to the regularized net extractable work $\frac{1}{n}W^{\varepsilon_1,\varepsilon_2}_{\beta}[\n^{\otimes n}\to\mc{M}^{\otimes n}]$, for $\varepsilon_1,\varepsilon_2\in(0,1)$,
\begin{align}
 \Delta W^\beta_{\rm ext}[\mc{N}\to \mc{M}]
 &\quad =\lim_{n\to\infty}\frac{1}{n}W^{\varepsilon_1,\varepsilon_2}_{\beta}[\n^{\otimes n}\to\mc{M}^{\otimes n}]\nonumber\\
    &\quad = \beta^{-1}\left(D[\n\|\mc{T}^\beta]-D[\mc{M}\|\mc{T}^\beta]\right)\nonumber\\
    &\quad = F^\beta_{\rm T}[\mc{N}]-F^\beta_{\rm T}[\mc{M}]\nonumber \\
    &\quad =W^\beta_{\rm ext}[\mc{N}]-W^{\beta}_{\rm ext}[\mc{M}].
\end{align}

The reversibility of the resource theory of athermality under GPSCs ensures that $\Delta W^\beta_{\rm ext}[\mc{N}\to \mc{M}]$ is asymptotically the optimal net extractable work in terms of the gain in athermality golden units during the thermodynamic transformation $\mc{N}\to\mc{M}$. It follows from the reversibility that $\Delta W^\beta_{\rm ext}[\mc{N}\to \mc{N}]=0$. The thermodynamic transformation is based on the distillation of $(\id_{m_1},\mc{R}^\pi)$ from $(\mc{N}^{\otimes n},{\mc{T}^{\beta}}^{\otimes n})$ and then expending $(\id_{m_2},\mc{R}^\pi)$ to form $(\mc{M}^{\otimes n},{\mc{T}^{\beta}}^{\otimes n})$, with largest possible $m_1$ and least possible $m_2$. Then the thermodynamic work capacity is given by $\Delta W^\beta_{\rm ext}[\mc{N}\to \mc{M}]= \lim_{n\to\infty}(\ln m_1-\ln m_2)/n=\lim_{n\to\infty}(1/n)\ln(m_1/m_2)$. Now, $\Delta W^\beta_{\rm ext}[\mc{N}\to \mc{M}]>0$ implies $m_1>m_2$, $\Delta W^\beta_{\rm ext}[\mc{N}\to \mc{M}]<0 $ implies $ m_1<m_2$, and $\Delta W^\beta_{\rm ext}[\mc{N}\to \mc{M}]=0$ implies $\lim_{n\to \infty}\frac{1}{n}\ln m_1=\lim_{n\to \infty} \frac{1}{n}\ln m_2$. 

\begin{table*}[htpb]
    \centering
    \renewcommand{\arraystretch}{1.0} 
    \textbf{Comparison between basic thermodynamic quantities of states and channels}\\[4pt]

    \setlength{\extrarowheight}{0pt}

    \begin{tabularx}{\textwidth}{
        >{\color{violet}\raggedright\arraybackslash}p{0.25\textwidth}
        >{\color{magenta}\raggedright\arraybackslash}p{0.34\textwidth}
        >{\color{blue}\raggedright\arraybackslash}p{0.36\textwidth}
    }
        \toprule
        \textbf{\vspace{-3pt}Concepts\vspace{-0.05pt}} & \textbf{\vspace{-3pt}States\vspace{-0.05pt}} & \textbf{\vspace{-3pt}Channels\vspace{-0.05pt}} \\
        \midrule
        Equilibration with bath & Thermal state $\gamma^\beta$  & Absolutely thermal channel $\mc{T}^\beta$\\
        Energy & $E(\rho)=\langle\widehat{H}\rangle_{\rho}$ & $E[\n]=\sup_{\rho\in\St(A')}\langle \widehat{H}\rangle_{\mc{N}(\rho)}$ \newline (for channel output noninteracting with its reference)\\
        Entropy & $S(\rho)=-\tr[\rho\ln\rho]=-D(\rho\|\mathbbm{1})$ & $S[\n]=-D[\n\Vert\mc{R}^{\mathbbm{1}}]$ \\
        Thermodynamic free energy & $F_{\rm T}^\beta(\rho)=\beta^{-1}D(\rho\|\widehat{\gamma}^\beta)$ & $F_{\rm T}^\beta[\n]=\beta^{-1}D[\n\|\widehat{\mc{T}}^\beta]$\\
        Resource-theoretic free energy & $F^\beta(\rho)=F^\beta_{\rm T}(\rho)-F^\beta_{\rm T}(\gamma^\beta)=\beta^{-1}D(\rho\Vert{\gamma}^\beta)$ & $F^\beta[\n]=F_{\rm T}^\beta[\n]-F_{\rm T}^\beta[\mc{T}^\beta]=\beta^{-1}D[\n\|{\mc{T}}^\beta]$ \\
     Minimal free energy & Thermal state $\gamma^\beta$ & Absolutely thermal channel $\mc{T}^\beta$\\
        Maximal extractable work & $W^\beta_{\rm ext}(\rho\to\gamma^\beta)=F^\beta(\rho)$ & $W^\beta_{\rm ext}[\mc{N}\to\mc{T}^\beta]=F^\beta[\n]$\\
       Helmholtz equation & $F^\beta_{\rm T}(\rho)=E(\rho)-\beta^{-1}S(\rho)$ & $F^{\beta}_{\rm T}[\n]\leq E[\n]-\beta^{-1}S[\n]$ \newline (for channel output noninteracting with its reference; inequality saturates for replacer channels)\\
        \bottomrule
    \end{tabularx}
    \caption{We summarize and compare mathematical expressions for elementary thermodynamic concepts for quantum states vs channels. We consider thermal reservoir to be at inverse temperature $\beta$ and $\widehat{H}$ denotes the Hamiltonian of the designated system. $\widehat{\gamma}^\beta$ is a thermal operator and $\gamma^\beta$ is a thermal state. $\mc{R}^\omega_{A'\to A}(\cdot):=\tr[\cdot]\omega_A$, $\mc{T}^\beta_{A'\to A}(\cdot):=\tr[\cdot]\gamma^\beta_A$, $\widehat{\mc{T}}^\beta_{A'\to A}(\cdot):=\tr[\cdot]\widehat{\gamma}^\beta$. $D(\rho\|\sigma):=\lim_{\varepsilon\to 0^+}\tr[\rho(\ln\rho-\ln(\sigma+\varepsilon\mathbbm{1}))]$ is the relative entropy between states $\rho,\sigma$ and $D[\mc{N}\|\mc{M}]:=\sup_{\psi\in\St(RA')}D(\id_R\otimes\mc{N}(\psi_{RA'})\|\id_R\otimes\mc{M}(\psi_{RA'}))$ is the relative entropy between channels $\mc{N}_{A'\to A},\mc{M}_{A'\to A}$. $\rho$ is a state of system $A'$ and $\mc{N}_{A'\to A}$ is a quantum channel. The content for quantum states are known in literature~\cite{BHN+15,BHO+13,DC19,Gou24}. The expression of the entropy of quantum channels is from \cite{GW21,SPSD25}. We provide proofs in the Appendix.}
    \label{tab:concepts}
\end{table*}
\textit{Discussion}.--- An arbitrary replacer channel $\mc{R}^{\omega}_{A'\to A}$ is an absolutely thermal channel $\mc{T}^{\beta}_{A'\to A}$ with respect to the channel output Hamiltonian $\widehat{H}_A=-\beta^{-1}\ln\omega_A$ and inverse temperature $\beta$. Our work provides the exact thermodynamic meaning to the generalized channel divergences like hypothesis-testing relative entropy, max-relative entropy, and relative entropy, between an arbitrary quantum channel $\n_{A'\to A}$ and a replacer channel $\mc{R}^{\omega}_{A'\to A}$, where $\omega_A$ is full rank and channel output Hamiltonian $\widehat{H}_A=-\beta^{-1}\ln \omega_A$; see for examples, Eqs.~\eqref{eq:w-ext}~and~\eqref{eq:distillation-formation}. Also, the resource theory of purity of states and channels~\cite{CG19,WW19,YZGZ20} can be seen as special cases of the resource theory of athermality of channels (see Table~\ref{tab:concepts} for a comparison of the thermodynamic concepts between states and channels).

Work is extractable from any athermal channel $\mc{N}$. This allows us to assess the utility of quantum gates (processors) as quantum batteries~\cite{CGQ+24,TSCG25,CDG+25} on the basis of their free energy. The physical transformation of a quantum channel $\n$ to $\mc{T}^\beta$ is the erasure of the channel $\n$ via thermalization~\cite{BGC+25} implying that the erasure cost of a channel is related to its free energy. We have detailed discussions on the quantitative relations between the free energy of a quantum channel and its private randomness distillation capacity~\cite{YHW19} and entropy~\cite{GW21,SPSD25} in Appendix~\ref{sec:aspects}. For instance, for a quantum channel $\n_{A'\to A}$ with $\widehat{H}_A\propto \mathbbm{1}_A$, its thermodynamic free energy is proportional to its entropy, $F_{\rm T}[\n]=-\beta^{-1}S[\n]$, thus providing an exact thermodynamic interpretation of the channel entropy~\cite{GW21,SPSD25,BGC+25}.

An interesting open question is to figure out if there exists an asymptotically reversible resource theory of athermality of quantum channels under a stricter and physically motivated class of free operations~(cf.~\cite{ST25}). Another future direction is to study the notion of free energy of higher-order quantum processes and to define the conditional free energy of bipartite quantum channels, and study their operational meanings in thermodynamic and information processing contexts (cf.~\cite{DGP24}). 

To define the free energy of higher order processes, one could take the approach in \cite[Sections V \& VI]{SPSD25} and define the inverse temperature times thermodynamic free energy of the $n$-th order quantum process ($n\in\mathbb{N}$, where $n=1$ for quantum states, $n=2$ for quantum channel, $n=3$ for superchannel and so on) as the relative entropy between the given $n$-th order quantum process and the absolutely thermal $n$-th order map. The absolutely thermal $n$-th order map on an input $(n-1)$-th order physical quantum process $\mc{M}^{(n-1)}$ is the replacer map $\mc{T}^{(n),\beta}(\mc{M}^{(n-1)})=\mc{T}^{(n-1),\beta}$ such that $\mc{T}^{(1),\beta}=\widehat{\gamma}^\beta$ and $\mc{T}^{(2),\beta}(\mc{M}^{(1)})=\mc{T}^{(1),\beta}$ for all input quantum states $\mc{M}^{(1)}$.

\begin{acknowledgements}
    The authors thank Felix C. Binder, Manabendra Nath Bera, Swati Choudhary, Karol Horodecki, Yutong Luo, and Uttam Singh for discussions. H.B. acknowledges support of the Visvesvaraya Post-Doctoral Fellowship (2025–26) in Electronics and IT from the Ministry of Electronics and Information Technology (MeitY), Govt.~of India. D.G.S. thanks the Department of Science and Technology (DST), Govt.~of India, for the INSPIRE fellowship. S.D. acknowledges support from the Science and Engineering Research Board (now ANRF), DST, Govt.~of India, under Grant No. SRG/2023/000217 and MeitY, Govt.~of India, under Grant No. 4(3)/2024-ITEA. S.D. also acknowledges support from the National Quantum Mission, an initiative of DST, Government of India.
\end{acknowledgements}

\onecolumngrid
\appendix
\section*{Appendix}
In this Appendix, we provide the elaborate proofs and discussions surrounding the main results in this article (see also \cite{BGD25} for related discussions from the perspective of an axiomatic resource-theoretic approach). We start by recalling the relevant definitions and notations from the literature in Section~\ref{sec:prem}. In Section~\ref{sec:prop}, we provide proofs for the expressions for the extractable work and the desirable properties exhibited by the resource-theoretic dynamical free energy $F^\beta[\mc{N}]$ claimed in this article. Along with this, we show that the unitary channels have the maximum free energy. We proceed to the detailed discussions on the resource theory of dynamical athermality in Section~\ref{sec:resourcetheory}, where we outline the motivation for defining $(\id,\mc{T}^\beta)$ as a golden unit of the resource theory. We determine the one-shot athermality distillation and formation of a quantum channel in Theorem~\ref{thm:dist_cost} and prove the asymptotic reversibility of the dynamical athermality under the action of Gibbs preserving superchannels in Theorem~\ref{thm:rev}. In  Section~\ref{sec:aspects} we present quantitative relations between information processing capabilities of a channel and its free energy, and finally in Section ~\ref{sec:energy} (Theorem~\ref{thm:fets}) we provide the Helmholtz equation for quantum channels.

\section{Preliminaries}\label{sec:prem}
We consider quantum systems associated with separable Hilbert spaces with tensor-product structure. For a bipartite state $\rho_{AB}$, $\rho_A=\tr_B(\rho_{AB})$. The energy of a quantum state $\rho_A$ associated with Hamiltonian $\widehat{H}_A$ is $\langle \widehat{H}\rangle_{\rho}:=\tr[\widehat{H}_A\rho_A]$. Let $\St(A)$ and $\Pos(A)$ denote the set of all density operators (quantum states) and the set of all positive semidefinite operators on $A$, respectively. Let $\Ch(A',A)$ denote the set of all quantum channels from $A'$ to $A$. Let $\SCh((A',A),(B',B))$ denote the set of all quantum superchannels mapping $\Ch(A',A)$ to $\Ch(B',B)$. Any quantum superchannel can be written as concatenation of preprocessing and postprocessing quantum channels~\cite{CDP09}, i.e., for any $\Theta\in\SCh((A',A),(B',B))$, there exists preprocessing channel $\mathcal{P}\in\Ch(B',RA')$ and postprocessing channel $\mathcal{Q}\in\Ch(RA,B)$ such that $\Theta(\mathcal{N})=\mathcal{Q}\circ\mathcal{N}\circ\mathcal{P}$. A unitary superchannel is composed of unitary postprocessing and preprocessing channels. Given two completely positive maps $\n$ and $\mc{M}$, we write $\n\ge \mc{M}$ if the linear map $\n-\mc{M}$  is also completely positive.

 $\id_R$ denotes the identity channel on $R$ and $\mathbbm{1}_R$ denotes the identity operator on $R$. Let $\Gamma_{RA}$ denote a maximally entangled operator, $\Gamma_{RA}:=\sum_{i,j=0}^{d-1}\ket{ii}\bra{jj}_{RA}$ for $d=\min\{|R|,|A|\}$; a maximally entangled state $\Phi_{RA}=\frac{1}{d}\Gamma_{RA}$ if $d<\infty$. For a linear supermap $\mc{N}_{A'\to A}$, $\Gamma^{\mc{N}}_{RA}=\id_R\otimes\mc{N}(\Gamma_{RA'})$ denotes its Choi operator. If $\mc{N}\in\Ch(A',A)$ is a finite-dimensional quantum channel, then $\Phi^{\mc{N}}_{RA}=\frac{1}{d}\Gamma^{\mc{N}}_{RA}$ is its Choi state. For $|A|<\infty$, $\pi_A:=\frac{1}{|A|}\mathbbm{1}_{A}$ is the maximally mixed state. $\norm{X}_p$ denotes the Schatten p-norm of an operator $X$. For any two Hermiticity-preserving maps $\mathcal{N},\mathcal{M}$, $\norm{\mc{N}-\mc{M}}_\diamond:=\sup_{\psi\in\St(RA')}\norm{\id_R\otimes\mc{N}(\psi_{RA'})-\id_R\otimes\mc{M}(\psi_{RA'})}_1$. The fidelity between $\rho,\sigma\in\St(A)$ is $F(\rho,\sigma):=\norm{\sqrt{\rho}\sqrt{\sigma}}_1^2$. Let $\langle\widehat{O}\rangle_{\rho_A}:=\tr[\widehat{O}_A\rho_A]$. For a quantum channel $\mc{N}_{A'\to A}$ and a bipartite state $\rho_{RA'}$, $\mc{N}(\rho_{RA'}):=\id_R\otimes\mc{N}(\rho_{RA'})$.

 \textit{Generalized state and channel divergences}.--- Let $\mathbf{D}(\cdot\Vert\cdot)$ be the generalized state divergence, i.e., for an arbitrary pair of quantum states $\rho,\sigma$ and an arbitrary quantum channel $\mathcal{N}$, we have $\mathbf{D}(\rho\Vert\sigma)\geq \mathbf{D}(\mathcal{N}(\rho)\Vert\mathcal{N}(\sigma))$. The generalized channel divergence $\bf{D}[\cdot\Vert\cdot]$ is derived from the generalized state divergence: For any two quantum channels $\mathcal{N},\mc{M}\in\Ch(A'A)$ and the identity channel $\id$,
\begin{equation}
    \mathbf{D}[\mathcal{N}\Vert\mathcal{M}]:=\sup_{\psi\in\mathrm{St}(RA')}\mathbf{D}(\id_R\otimes\mathcal{N}(\psi)\Vert\id_R\otimes\mathcal{M}(\psi)),
\end{equation}
where it suffices to optimize over pure states $\psi_{RA'}$~\cite{LKDW18}. Some examples of the generalized state divergences~\cite{Tom21,WWY14,Dat09,LR12}: For a state $\rho_A$ and a positive semidefinite operator $\sigma_A$, whenever $\operatorname{supp}(\rho) \subseteq \operatorname{supp}(\sigma)$ (otherwise $+\infty$) (a) the quantum relative entropy $D(\rho\Vert\sigma):=\operatorname{tr}\left[\rho (\ln \rho - \ln \sigma)\right]$, (b) the max-relative entropy $D_{\infty}(\rho\Vert\sigma):=\ln \norm{\sigma^{-1/2} \rho \, \sigma^{-1/2}}_{\infty}$, (c) the sandwiched R\'enyi relative entropy ${D}_\alpha(\rho \Vert \sigma) := \frac{1}{\alpha - 1} \ln \left( \operatorname{Tr} \left[ \left( \sigma^{\frac{1 - \alpha}{2\alpha}} \rho \, \sigma^{\frac{1 - \alpha}{2\alpha}} \right)^\alpha \right] \right)$ for $\alpha\in[\frac{1}{2},1)\cup(1,\infty)$, (d) the $\varepsilon$-hypothesis-testing relative entropy $D^{\varepsilon}_{\mathrm{H}}(\rho\Vert\sigma):=-\ln \inf_{0 \leq \Lambda \leq \mathbbm{1}} \left\{ \operatorname{tr}[\Lambda \sigma] \, : \, \operatorname{tr}[\Lambda \rho] \geq 1 - \varepsilon \right\}$, (e) the purified distance $\mathrm{P}(\rho,\sigma):=\sqrt{1-F(\rho,\sigma)}$. $\ln$ denotes natural logarithm. We have $\lim_{\alpha\to 1}D_{\alpha}(\rho\|\sigma)=D(\rho\|\sigma)$ and $\lim_{\alpha\to \infty}D_{\alpha}(\rho\|\sigma)=D_{\infty}(\rho\|\sigma)$. 

The generalized channel divergences are monotonically nonincreasing under the action of quantum superchannels~\cite[Theorem 1]{DGP24}. If the generalized state divergence is nonnegative, then the corresponding channel divergence is also nonnegative. If the generalized state divergence is faithful, i.e., minimum if and only the states are equal, then the corresponding channel divergence is also faithful, i.e., attains minimum if and only if the channels are equal.

An $\varepsilon$-ball around a state $\rho_A$, for $\varepsilon\in[0,1]$, is defined as $\mathcal{B}^\varepsilon(\rho_A)=\{\sigma_A:~ \sigma_A\ge 0, \tr(\sigma_A)\le 1, P(\rho_A,\sigma_A)\le \varepsilon\}$. For $\rho\in\St(A)$ and $\sigma\in\Pos(A)$, $D_{\infty}^\epsilon(\rho\|\sigma) = \inf_{\omega\in\mc{B}^{\varepsilon}(\rho)} D_{\infty}(\omega\|\sigma)$. An $\varepsilon$-ball around a channel $\mc{N}_{A'\to A}$ is defined as $\mathcal{B}^\varepsilon[\mc{N}]=\{\mc{M}\in\Ch(A',A): P[\mc{N},\mc{M}]\leq \varepsilon\}$. We use this to define smoothed channel max-relative entropy as $D_{\infty}^{\varepsilon}[\n\|\mc{M}]=\inf_{\mathcal{Q}\in \mathcal{B}^\varepsilon[\mc{N}]}D_{\infty}[\mathcal{Q}\|\mc{M}]$. $\textbf{D}^\varepsilon(\cdot\Vert\cdot)$ is a generalized smoothed-state divergence and $\textbf{D}^{\varepsilon}[\cdot\|\cdot]$ is a generalized smoothed-channel divergence.

The channel entropy $S[\n]$ of a quantum channel $\n_{A'\to A}$ is defined as $S[\n]:=-D[\mc{N}\Vert\mc{R}^{\mathbbm{1}}]$~\cite{GW21,SPSD25}, where a completely positive map $\mc{R}^{\omega}_{A'\to A}$ denotes the replacer (or preparation) map that outputs $\omega\in\Pos(A)$, $\mc{R}^{\omega}_{A'\to A}(\rho_{A'})=\tr[\rho_{A'}]\omega_A$ for all $\rho_{A'}\in\Pos(A')$. The (von Neumann) entropy of a quantum state $\rho_A$ is $S(\rho_A):=S(A)_{\rho}:=-D(\rho\|\mathbbm{1})=-\tr[\rho\ln\rho]$. $\mc{R}^\pi$ is called a uniformly mixing or completely depolarizing channel, as it always outputs the maximally mixed state $\pi (\propto \mathbbm{1})$.
\section{Extractable work and dynamical free energy}\label{sec:prop}
For a fixed inverse temperature $\beta$, the thermodynamic non-equilibrium free energy $F^\beta_{\rm T}(\rho)$ of a quantum state $\rho_A$ is related to its entropy and energy as
\begin{equation}
    F^\beta_{\rm T}(\rho)=E(\rho)-\beta^{-1}S(\rho)=\beta^{-1}D(\rho\Vert\widehat{\gamma}^{\beta}),
\end{equation}
where $\widehat{\gamma}^{\beta}_A=\exp(-\beta \widehat{H}_A)$ is the thermal operator (unnormalized thermal state).
\begin{lemma}
For a quantum state $\rho_{AB}$ with the Hamiltonian $\widehat{H}_{AB}=\mathbb{1}_A\otimes\widehat{H}_B+\widehat{H}_A\otimes\mathbb{1}_B$ that is weakly interacting with the bath at temperature $\beta$, we have 
\begin{align}
 F^\beta_{\rm T}(\rho_{AB})-F^\beta_{\rm T}(\gamma^\beta_A\otimes\gamma^\beta_B)=F^\beta(\rho_{AB})=\beta^{-1}I(A;B)_{\rho}+F^\beta(\rho_A)+F^\beta(\rho_B).
\end{align}
\end{lemma}
\begin{proof}
Recall that $F_T^\beta(\rho)=E(\rho)+\beta^{-1}S(\rho)$ where $E(\rho)=\tr(\rho \widehat{H})$ is the energy of the system. For noninteracting Hamiltonian, we have $E(\rho_{AB})=E(\rho_A)+E(\rho_B)$. Furthermore, since $S(\gamma^\beta_A\otimes\gamma_B^\beta)=S(\gamma_A^\beta)+S(\gamma_B^\beta)$ we have $F_T^\beta(\gamma^\beta_A\otimes\gamma_B^\beta)=F_T^\beta(\gamma^\beta_A)+F_T^\beta(\gamma_B^\beta)$, which gives
\begin{align}
    F_T^\beta(\rho_{AB})-F_T^\beta(\gamma^\beta_A\otimes\gamma_B^\beta)&=E(\rho_A)+E(\rho_B)-\beta^{-1}S(\rho_{AB})-F_T^\beta(\gamma^\beta_A)-F_T^\beta(\gamma_B^\beta)\\
    &=E(\rho_A)-\beta^{-1}S(\rho_A)+E(\rho_B)-\beta^{-1}S(\rho_B)+\beta^{-1}S(\rho_A)+\beta^{-1}S(\rho_B)-\beta^{-1}S(\rho_{AB})\nonumber\\&\qquad-F_T^\beta(\gamma_A^\beta)-F^\beta_T(\gamma_B^\beta)\\
&=F_T^\beta(\rho_A)+F_T^\beta(\rho_B)+\beta^{-1}S(\rho_A)+\beta^{-1}S(\rho_B)-\beta^{-1}S(\rho_{AB})-F_T^\beta(\gamma_A^\beta)-F^\beta_T(\gamma_B^\beta)\\
&=F^\beta(\rho_A)+F^\beta(\rho_B)+\beta^{-1}I(A;B)_\rho.
\end{align}
\end{proof}

\begin{lemma}
    The extractable work $W^{\beta}_{\rm ext}[\mc{N}]$ from a quantum channel $\mc{N}_{A'\to A}$ is equal to
    \begin{equation}\label{eq:chan-work}
        W^{\beta}_{\rm ext}[\mc{N}]=\sup_{\psi\in\St(RA')}\left[\beta^{-1}I(R;A)_{\id\otimes\n(\psi)}+F^\beta(\n(\psi_{A'}))\right]=\beta^{-1}D[\mc{N}\|\mc{T}^\beta],
    \end{equation}
    where $\mc{T}^\beta_{A'\to A}$ is the absolutely thermal channel, $\mc{T}^\beta(\cdot)=\tr(\cdot)\gamma^\beta_A$.
\end{lemma}
\begin{proof}
It follows from the definition of the extractable work from a channel $\n_{A'\to A}$ that
\begin{align}
    W^{\beta}_{\rm ext}[\n]
    &:=\sup_{\psi\in\St(RA')}\left[F^\beta_{\rm T}(\id\otimes\mc{N}(\psi))-F^\beta_{\rm T}(\psi_R\otimes\gamma^\beta_A)\right]\nonumber\\
    &=\sup_{\psi\in\St(RA')}\left[F^\beta(\id\otimes\mc{N}(\psi))-F^\beta(\psi_R\otimes\gamma^\beta_A)\right]\nonumber\\
&=\sup_{\psi\in\St(RA')}\left[\beta^{-1}I(R;A)_{\id\otimes\n(\psi)}+F^\beta(\n(\psi_{A'}))\right].
\end{align}
We know that $F^\beta(\n(\psi_{A'}))=\beta^{-1}D(\n(\psi_{A'})\Vert \gamma_A^\beta)$ and $I(A;B)_{\rho}+D(\rho_B\Vert\gamma^\beta)=D(\rho_{AB}\|\rho_A\otimes\gamma^\beta_B)$, which implies 
\begin{align}
      W^{\beta}_{\rm ext}[\n]
&=\sup_{\psi\in\St(RA')}\left[\beta^{-1}I(R;A)_{\id\otimes\n(\psi)}+F^\beta(\n(\psi_{A'}))\right]\\
&=\sup_{\psi\in\St(RA')}\left[ \beta^{-1}D(\id\otimes\n(\psi)\Vert \id\otimes \mc{T}^\beta(\psi))\right]\\
&=\beta^{-1}D[\n\Vert\mc{T}^\beta].
\end{align}
\end{proof}

\begin{proposition}[\cite{BGD25}]\label{prop:free-eenrgy-properties}
    The (resource-theoretic) free energy $F^\beta[\n]:=\beta^{-1}D[\n_{A' \to A} \| \mathcal{T}^{\beta}_{A' \to A}] $ of an arbitrary quantum channel $\mc{N}_{A'\to A}$ satisfies the following properties:
     \begin{enumerate}
     \item[(P1)] Monotonically nonincreasing under the action of a Gibbs preserving superchannel $\Theta^{\beta}$, $\Theta^\beta(\mathcal{T}^{\beta}_{A'\to A})=\mathcal{T}^{\beta}_{B'\to B}$, $F^\beta[\Theta^\beta(\mathcal{N})]\leq F^\beta[\mathcal{N}]$.
        \item[(P2)] Reduction to the free energy function of the output state for a replacer channel $\mathcal{R}^{\omega}$, $\mathcal{R}^{\omega}(\rho_{A'}):=\omega\in\St(A)$ for all input states $\rho_{A'}$, $F^\beta[\mathcal{R}^{\omega}]=F^\beta(\omega)$.
     \item[(P3)] Minimum is attained if and only if the channel is absolutely thermal.
      \item[(P4)] Continuity: For any two channels $\mathcal{N},\mathcal{M}$ such that $\frac{1}{2}\norm{\mathcal{N}-\mathcal{M}}_{\diamond}\leq \varepsilon\in[0,1]$, we have $\abs{F^\beta[\mathcal{N}]-F^\beta[{\mc{M}}]}\le {\beta}^{-1} (\varepsilon K + h_2(\varepsilon))$, where, $K$ is a positive number given in Eq.~\eqref{eq:K}.
     \item[(P5)] Additive under tensor-product channels $\mathcal{N}_{A_1'\to A_1}\otimes\mathcal{M}_{A_2'\to A_2}$ with $\widehat{H}_{A_1A_2}=\widehat{H}_{A_1}\otimes\mathbbm{1}_{A_2}+\mathbbm{1}_{A_1}\otimes\widehat{H}_{A_2}$, $F^\beta[\mathcal{N}\otimes\mathcal{M}]=F^\beta[\mathcal{N}]+F^\beta[\mathcal{M}]$.
     \item[(P6)] Convexity: $F^\beta[p\mathcal{N}+(1-p)\mathcal{M}]\leq pF^\beta[\mathcal{N}]+(1-p)F^\beta[\mathcal{M}]$ for $p\in[0,1]$, where $\mathcal{N}_{A'\to A}$ and $\mathcal{M}_{A'\to A}$ are quantum channels.
 \end{enumerate}
\end{proposition}
 
\begin{proof}
\begin{enumerate}
\item[(P1)] \textit{Monotonicity under Gibbs preserving superchannels}: This follows from the monotonicity of relative entropy between channels. Let $\Theta$ be a superchannel such that $\Theta(\mathcal{T}^\beta_{A'\to A})=\mathcal{T}^\beta_{B'\to B}$, it follows that 
\begin{align}
D[\n_{A' \to A} \| \mathcal{T}^{\beta}_{A' \to A}] & \geq D[\Theta (\n_{A' \to A}) \| \Theta( \mathcal{T}^{\beta}_{A' \to A})]\\
    &= D[\Theta (\n_{A' \to A}) \|  \mathcal{T}^{\beta}_{A' \to A}],
\end{align}
which implies $ F^\beta[\Theta (\n_{A' \to A})] \le F^\beta[\n_{A' \to A}]$.
\item[(P2)] \textit{Reduction to states}:  For a replacer channel $\mathcal{R}^{\omega}_{A'\to A}(\rho_{A'}):=\tr(\rho_{A'})\omega_A$, we have
\begin{align}
            \beta F^\beta[\mathcal{R}^{\omega}]&=D[\mathcal{R}^\omega\Vert\mathcal{T}^{\beta}]\nonumber\\
            &= \sup_{\psi\in\St(RA')}D(\mathcal{R}^{\omega}(\psi_{RA'})\Vert\mathcal{T}^\beta(\psi_{RA'}))\nonumber\\
            &=\sup_{\psi\in\St(RA')}D(\psi_R\otimes\omega_A\Vert\psi_R\otimes\gamma^\beta_A)\nonumber\\
            &= D(\omega_A\Vert\gamma^\beta_A)=\beta F^\beta(\omega).
        \end{align}
\item[(P3)] \textit{Minimum value at absolutely thermal channel}: This follows from the fact that $D(\rho\Vert\sigma)\ge 0$ if both $\rho$ and $\sigma$ are states, and $D(\rho\|\sigma)= 0$ if and only if $\rho=\sigma$. Therefore, given quantum channels $\n_{A'\to A}$ and $\mc{M}_{A'\to A}$, $D[\mc{N}\Vert\mc{M}]=\sup_{\psi\in\St(RA')} D(\n(\psi)\|\mathcal{M}(\psi))=0$ if and only if $\n(\varphi)=\mathcal{M}(\varphi)$ for all $\varphi\in \St(RA')$ or $\n=\mathcal{M}$. Therefore, $F^\beta[\n]=\beta ^{-1} D[\n\|\mathcal{T}^\beta]\ge 0$ with equality if and only if $\n=\mathcal{T}^\beta$.
    \item[(P4)] \textit{Continuity:} We shall show that given two quantum channels $\n_{A' \to A}$ and $\mathcal{M}_{A' \to A}$ such that $\frac{1}{2}\|\n-\mathcal{M}\|_{\diamond} \le \varepsilon$, we have
    \begin{align} \label{eq:continuity}
        \abs{F^\beta[\n] -  F^\beta[\mathcal{M}]} \le {\beta}^{-1} (\varepsilon K + h_2(\varepsilon)),
        \end{align}
where $\varepsilon \in [0,1]$ and $h_2(\varepsilon) =-\varepsilon\ln(\varepsilon)-(1-\varepsilon)\ln(1-\varepsilon)$ is the binary Shannon entropy. $K$ is a positive real number such that
        \begin{equation}\label{eq:K}
       K\ge \max \{ D_\infty(\n(\psi^0_{RA'})\|\mathcal{T}^\beta(\psi^0_{RA'})),D_\infty(\mc{M}(\phi^0_{RA'})\Vert\mc{T}^\beta(\phi^0_{RA'}))\},
        \end{equation}
where $\psi^0_{RA'}$ maximizes $D(\n(\psi_{RA'})\|\mathcal{T}^\beta(\psi_{RA'}))$ and $\phi^0_{RA'}$ maximizes $D(\mc{M}(\psi_{RA'})\|\mathcal{T}^\beta(\psi_{RA'}))$.

This can be shown by noting that the condition $\frac{1}{2}\|\n-\mathcal{M}\|_{\diamond} \le \varepsilon$ implies that $\frac{1}{2}\|\n(\psi_{RA'})-\mathcal{M}(\psi_{RA'})\|_1 \le \varepsilon$ for all $\psi_{RA'}$. The continuity of relative entropy in the first argument has been tightened in \cite{BLT25,ABD+25}. Here we use a slightly simpler bound given in \cite[Theorem 1]{BLT25}, from which it follows that
    \begin{align}
        D(\n(\psi_{RA'})\|\mathcal{T}^\beta(\psi_{RA'})) - D(\mathcal{M}(\psi_{RA'})\|\mathcal{T}^\beta(\psi_{RA'})) 
        \le \varepsilon K + h_2(\varepsilon),
    \end{align}
where, $  K\ge D_\infty(\n(\psi_{RA'}) \| \mathcal{T}^\beta(\psi_{RA'}))$ is some positive real number. Therefore, we have
\begin{align}\label{eq:n-m_continuity}
    D[\n\|\mathcal{T}^\beta] - D[\mathcal{M}\|\mathcal{T}^\beta]\le D(\n(\psi^0_{RA'})\|\mathcal{T}^\beta(\psi^0_{RA'})) - D(\mathcal{M}(\psi^0_{RA'})\|\mathcal{T}^\beta(\psi^0_{RA'})) \le \varepsilon K_{\n} + h_2(\varepsilon), 
\end{align}
where the state $\psi^0_{RA'}$ maximizes $D(\n(\psi_{RA'})\|\mathcal{T}^\beta(\psi_{RA'}))$ and $K_{\n}\ge D_\infty(\n(\psi^0_{RA'})\|\mathcal{T}^\beta(\psi^0_{RA'}))$. Following the similar steps for the channel $\mathcal{M}$,
\begin{align}\label{eq:m-n_continuity}
    D[\mathcal{M}\|\mathcal{T}^\beta] - D[\n\|\mathcal{T}^\beta]
    \le  \varepsilon  K_{\mc{M}} + h_2(\varepsilon),
\end{align}
where $K_{\mc{M}}\ge D_\infty(\mc{M}(\phi^0_{RA'})\Vert\mc{T}^\beta(\phi^0_{RA'}))$ where the state $\phi^0_{RA'}$ maximizes $D(\mc{M}(\psi_{RA'})\|\mathcal{T}^\beta(\psi_{RA'}))$. From Eqs.~\eqref{eq:n-m_continuity} and~\eqref{eq:m-n_continuity}, we have
\begin{align}
    &|D[\n\|\mathcal{T}^\beta] - D[\mathcal{M}\|\mathcal{T}^\beta]| \le  \varepsilon K + h_2(\varepsilon),
\end{align}
where $K=\max\{K_{\n},K_{\mc{M}}\}$. The continuity relation \eqref{eq:continuity} follows by using the definition of free energy $F^\beta[\n]=\beta^{-1} D[\n\Vert\mc{T}^\beta]$ in the above inequality.
\item[(P5)] \textit{Additivity}: For two channels $\n_{A_1' \to A_1}$ and $\mc{M}_{A_2' \to A_2}$ such that the output interaction Hamiltonian $\widehat{H}_{A_1A_2}$ is zero, the absolutely thermal channel can be written as $\mc{T}^\beta_{A_1'A_2'\to A_1A_2}=\mc{T}^\beta_{A_1' \to A_1} \otimes \mc{T}^\beta_{A_2' \to A_2}=:\mc{T}^\beta_1 \otimes \mc{T}^\beta_2$.  In this case, the additivity of the generalized free energy of channel follows from the additivity of the relative channel entropy since
\begin{equation}
    \beta F^\beta[\n \otimes \mc{M}] = D[\n \otimes \mc{M}\|\mathcal{T}^\beta_1 \otimes \mathcal{T}^\beta_2].
\end{equation}
We will show that the sandwiched R\'enyi channel divergence is additive under tensor product with respect to the absolutely thermal channel, i.e. 
\begin{align}
    D[\n \otimes \mc{M}\|\mathcal{T}^\beta_1 \otimes \mathcal{T}^\beta_2]= D[\n\|\mathcal{T}^\beta_1] + D[\mc{M}\| \mathcal{T}^\beta_2].
\end{align}
The proof of the inequality "$\ge$" follows from the definition of the  channel relative entropy,
\begin{align}
    D[\n \otimes \mc{M}\|\mathcal{T}^\beta_1 \otimes \mathcal{T}^\beta_2] \nonumber &= \sup_{\psi\in\St(RA_1'A_2')} D(\n \otimes \mc{M} (\psi)\| \mathcal{T}^\beta_1 \otimes \mathcal{T}^\beta_2(\psi))\\
    & \ge  \sup_{\varphi\in\St(R_1A_1')\otimes\St(R_2A_2')} D(\n \otimes \mc{M} (\varphi)\| \mathcal{T}^\beta_1 \otimes \mathcal{T}^\beta_2(\varphi)) \\
    &= D[\n\|\mathcal{T}^\beta_1] + D[\mc{M}\| \mathcal{T}^\beta_2],
\end{align}
where we have used the additivity of the relative entropy of states under the tensor product. 

To prove the inequality ``$\le$", we use the following result from \cite[Lemma 38]{WBHK20}, for an arbitrary state $\rho_{RA'}$, a channel $\n_{A'\to A}$ , and a replacer channel $\mc{R}^{\tau}_{A'\to A}$ which replaces all its input states with a state $\tau$,
\begin{align}
D(\n(\rho_{RA'})\|\sigma_{R} \otimes \tau_A) \le D[\n\|\mathcal{R}^\tau] + D(\rho_{RA'}\|\sigma_{RA'}).
\end{align}
Let $\psi_{RA_1'A_2'}$ be an arbitrary pure state and states $ \rho_{R'A_1'}$ and $\sigma_{R'A_1'}$ be such that $R'=RA_2$, i.e., $\mc{M}_{A_2' \to A_2}(\psi_{RA_1'A_2'})=\rho_{R'A_1'}$ and $\mathcal{T}^\beta_{A_2' \to A_2}(\psi_{RA_1'A_2'})=\sigma_{R'A_1'}$,
\begin{align}
    D(\n \otimes \mc{M} (\psi_{RA_1'A_2'})\| \mathcal{T}^\beta_{A_1' \to A_1} \otimes \mathcal{T}^\beta_{A_2' \to A_2}(\psi_{RA_1'A_2'})) &\le D[\n\|\mathcal{T}^\beta] + D(\mc{M}(\psi_{RA_1'A_2'}) \| \mathcal{T}^\beta_{A_2' \to A_2}(\psi_{RA_1'A_2'}))\\
    & \le D[\n\|\mathcal{T}^\beta] + D[\mc{M}\|\mathcal{T}^\beta].
\end{align}
This implies the additivity of  $F^\beta[\n]$.
\item[(P6)]  \textit{Convexity}: It is known that quantum relative entropy is jointly convex, that is, for a probability distribution $\{p_x\}_x$, we have 
\begin{align}
  D\left[\sum_x p_x \n^x\|\mathcal{T}^\beta\right]
    & = \sup_{\psi_{RA'}} D\left(\sum_x p_x \n^x(\psi)\| \mathcal{T}^\beta(\psi)\right)\\
    & \le \sup_{\psi_{RA'}} \sum_x p_x D( \n^x(\psi)\| \mathcal{T}^\beta(\psi)),
\end{align}
which implies
\begin{align}
F^\beta\left[\sum_xp_x\n^x\right]\le\sum_xp_xF^\beta[\n^x].
\end{align}
\end{enumerate}
\end{proof}
\subsection{Free energy for unitary channels}\label{app:proof-thm-unitary-free-energy}
Before we show that for a given bath, unitary channels have the maximum channel resource, we prove the following lemma.
\begin{lemma}\label{lemma:convex_purestate}
    Given a state $\rho=\sum_ip_i\op{\psi_i}$, the max-relative entropy is bounded as follows
    \begin{align}
D_\infty(\rho\Vert\sigma)\ge \ln(\max_ip_i\bra{\psi_i}\sigma^{-1}\ket{\psi_i}),\\
D_\infty(\rho\Vert\sigma)\le\ln(\sum_ip_i\bra{\psi_i}\sigma^{-1}\ket{\psi_i}).
    \end{align}
\end{lemma}
\begin{proof}
The max-relative entropy $D_\infty(\rho\Vert\sigma)$ can be written as
    \begin{align}
        D_\infty(\rho\Vert\sigma)&=\ln\Vert\sum_ip_i\sigma^{-\frac{1}{2}}\ket{\psi_i}\bra{\psi_i}\sigma^{-\frac{1}{2}}\Vert_\infty\\
        &=\ln\Vert AA^\dagger\Vert_\infty\\
        &=\ln\lambda_{\max}(AA^\dagger),
    \end{align}
 where the matrix $A$ is given by $A=[\ket{v_1}~~\ket{v_2}~...]$ with the vectors $\ket{v_i}:=\sqrt{p_i}\sigma^{-\frac{1}{2}}\ket{\psi_i}$. The corresponding Gram matrix of matrix $A$ is given by $G=A^\dagger A$, such that $G_{ij}=\braket{v_i}{v_j}=\sqrt{p_ip_j}\bra{\psi_i}\sigma^{-1}\ket{\psi_j}$. Note that $AA^\dagger$ and $G$ have the same non-zero eigenvalues. Therefore, $D_\infty(\rho\Vert\sigma)=\ln(\lambda_{\max}(AA^\dagger))=\ln(\lambda_{\max}(G))$. Now, for any matrix $G$ we have the following property
 \begin{align}
     \max_{i}(G_{ii})\le\lambda_{\max}(G)\le \tr(G).
 \end{align}
 Using the fact that $G_{ij}=\braket{v_i}{v_j}=\sqrt{p_ip_j}\bra{\psi_i}\sigma^{-1}\ket{\psi_j}$, we have the proof of the stated lemma.
\end{proof}
\begin{proposition}
For $\alpha\in\{1,\infty\}$ and $|A'|=|A|$, the free energy $F^{\beta}_{\alpha}[\mc{N}]$ of a quantum channel $\n_{A'\to A}$ is maximum if and only if $\n$ is a unitary channel. For all unitary channels $\mc{U}_{A'\to A}$, $F^{\beta}_{\alpha}[\mc{U}]=F^{\beta}_{\alpha}[\id_{A'\to A}]$.
\end{proposition}
\begin{proof}
Given a state $\rho_{RA}$ and its decomposition into pure states $\rho_{RA}=\sum_xp_x\psi^x_{RA}$, we use the joint quasi-convexity of the sandwiched R\'enyi entropy to observe that for $\alpha\in\{1,\infty\}$
\begin{align}\label{eq:joint-quasi-convex}
D_\alpha(\sum_xp_x\psi^x_{RA}\Vert\sum_xp_x\rho^x_{R}\otimes\gamma^\beta_A)\le \max_x D_\alpha(\psi^x_{RA}\Vert\rho^x_{R}\otimes\gamma^\beta_A),
\end{align}
where $\rho^x_{R}=\tr_A\psi_{RA}^x$.  Therefore, $D_{\alpha}(\rho_{RA}\Vert\rho_R\otimes\gamma_A)$ achieves the maximum value if $\rho_{RA}$ is a pure state. This implies that $D_\alpha[\n\Vert\mc{T}^\beta]$ achieves the maximum value if $\n$ is a unitary channel.

We will show that the ``only if" condition is satisfied for $\alpha\in\{1,\infty\}$ by proving that the equality in Eq.~\eqref{eq:joint-quasi-convex} holds if and only if $\rho_{RA}$ is a pure state. Since we know that relative entropy is jointly convex, and therefore for $\rho=\sum_i p_i\psi_i$, where $\{\psi_i\}$ are pure, we have   
\begin{align}
    D(\rho\Vert\sigma)\le\sum_ip_iD(\psi_i\Vert\sigma).
\end{align}
The equality in the above holds if and only if $\rho$ is pure~\cite{Rus02}.

For the case of $\alpha= \infty$, let the equality for the quasi-convexity condition be satisfied, i.e.  $D_\infty(\rho\Vert\sigma)=\max_iD_\infty(\psi_i\Vert\sigma)$, then using Lemma~\ref{lemma:convex_purestate},
\begin{align}
  \max_i \ln(\bra{\psi_i}\sigma^{-1}\ket{\psi_i})&\le\ln(\sum_ip_i\bra{\psi_i}\sigma^{-1}\ket{\psi_i})\\
\text{or}~  \max_i\bra{\psi_i}\sigma^{-1}\ket{\psi_i}&\le\sum_ip_i\bra{\psi_i}\sigma^{-1}\ket{\psi_i}.
\end{align}
Since the weighted average cannot be greater than the largest element of the set, the above condition is   satisfied if and only if $\rho$ is a pure state.
\end{proof}
\section{Resource theory of dynamical athermality}\label{sec:resourcetheory}
We now introduce framework and characterization of the thermodynamic resource theory of square quantum channels. A quantum channel $\n_{A'\to A}$ is a square quantum channel if $|A'|=|A|$; in other words, a quantum channel with square Choi operator is called a square channel. To quantify the thermodynamic resourcefulness of a quantum channel $\n$, we consider its distinguishability from a thermal channel $\mc{T}^{\beta}$ whose output $\gamma^\beta$ is in thermal equilibrium with the bath~\cite{DS25}. The absolutely thermal channel $\mc{T}^{\beta}$ represents thermalization process with respect to the given bath at inverse temperature $\beta$. 

To characterize a resource theory of athermality of quantum channels, we identify our free object to be the absolutely thermal channel $\mc{T}^\beta$ and the free operations to be Gibbs preserving superchannels (GPSC). This is a relevant assumption as thermalization processes seem to be natural in practical scenarios where quantum systems inherently tend to thermalize upon interaction with the bath. The closed systems in isolation with the bath are nontrivial and idealistic. We write the athermal channel resource as $(\n,\mathcal{T}^\beta)$ to specify that the athermality of the channel $\n$ is being considered with respect to $\mc{T}^{\beta}$. We call two channel-resources $(\n,\mathcal{T}^\beta)$ and $(\mathcal{M},\mathcal{T}^{\beta'})$ to be equivalent, $(\n,\mathcal{T}^\beta)\sim(\mathcal{M},\mathcal{T}^{\beta'})$, if they are interconvertible through some GPSCs.

For a quantum superchannel $\Theta\in\SCh((A',A),(B',B))$ with the following action,
\begin{equation}
    \Theta(\n_{A'\to A})=\mc{Q}_{A\to B}\circ\n\circ\mc{P}_{B'\to A'},
\end{equation}
$\Theta$ is Gibbs preserving if and only if $\mc{Q}_{A\to B}$ is Gibbs preserving channel, i.e., $\mc{Q}(\gamma^\beta_A)=\gamma^{\beta}_B$. A quantum superchannel $\widetilde{\Theta}\in\SCh((A',A),(B',B))$, where $\widetilde{\Theta}(\n)=\widetilde{\mc{Q}}_{PA\to B}\circ\n\circ\widetilde{\mc{P}}_{B'\to PA'}$ is also Gibbs preserving if its postprocessing channel $\widetilde{\mc{Q}}_{PA\to B}$ is absolutely thermal. 

 \subsection{Golden unit of the dynamical athermality}\label{subsec:goldenunit} 
 For a square quantum channel $\n_{A'\to A}$, the free energy is maximum if and only if the channel is unitary. For an arbitrary pair $\mc{U}_{A'\to A},\mc{V}_{A'\to A}$ of unitary channels, there always exists a GPSC $\Theta^{\beta}\in\SCh((A',A),(A',A))$ such that $\Theta^{\beta}(\mc{U})=\mc{V}$, for an instance, $\Theta^{\beta}(\n)=\n\circ(\mc{U}^{\dag}\circ\mc{V})$. That is, all unitary channels $(\mc{U},\mc{T}^\beta)$ are equally resourceful, $(\id_{A'\to A},\mc{T}^\beta_{A'\to A})\sim(\mc{U}_{A'\to A},\mc{T}^\beta_{A'\to A})$. This is consistent with the free energy of all unitary channels $(\mc{U},\mc{T}^\beta)$ being the same. We can show that the free energy of a unitary channel is minimum when the output Hamiltonian is fully degenerate. To see this, we will use the following Lemmas.
\begin{lemma}[\cite{LMB25}]\label{prop:unitary_free_energy}
    The max-free energy $F^{\beta}_{\infty}[\mc{U}]$ of a unitary quantum channel $\mc{U}_{A'\to A}$ is
    \begin{equation}
        F^{\beta}_{\infty}[\mc{U}]= \beta^{-1}\ln \tr\left[{(\gamma^{\beta}_A)}^{-1}\right].
    \end{equation}
    If the Hamiltonian of $A$ is fully degenerate, $\widehat{H}_A\propto \mathbbm{1}_A$, then $F^{\beta}_{\infty}[\mc{U}]=2\beta^{-1}\ln|A|$.
\end{lemma}
The proof follows from \cite[Corollary A.4.2]{LMB25}, since $F_{\infty}^\beta[\n]=\beta^{-1}\ln(1+R^\beta_T[\n])$. For the sake of completeness, we give a proof.
\begin{proof}
Using the property of the max-relative entropy of quantum channels
\begin{equation}\label{eq:max-relequivalence}
\beta F_{\infty}^\beta [\mc{U}] = D_{\infty}[\mathcal{U} \| \mathcal{T}^{\beta}] = D_{\infty}(\Phi^{\mathcal{U}} \|\Phi^{\mathcal{T}^\beta}),
\end{equation}
where, $\Phi^\mc{U}$ is a pure state and $\Phi^{\mc{T}^\beta}=\pi_R\otimes\gamma_{\beta}$. Given a pure state $\psi$, we have $D_\infty(\psi\Vert\sigma)=\ln\bra{\psi}\sigma^{-1}\ket{\psi}$ for a positive operator $\sigma$. Let $ \mathcal{U}(\rho) = U \rho U^\dagger$, then
\begin{align}
\beta F_{\infty}^\beta [\mc{U}]  &= D_{\infty}(\Phi^{\mathcal{U}} \|\Phi^\mathcal{T}) \\
 &= \ln \left(\frac{1}{d} \sum_{j} 
(\bra{j} \otimes \bra{j} U^\dagger) 
\left( \frac{1}{d} \mathbbm{1}_R\otimes \gamma_\beta \right)^{-1} \right. \nonumber  \left. \sum_{i} (\ket{i} \otimes U \ket{i}) \right) \nonumber \\
&= \ln \left( \sum_{i,j} 
\delta_{ji}  \bra{j} U^\dagger \gamma_\beta^{-1} U \ket{i} \right) \nonumber \\
&= \ln \left( \tr \left[ \gamma_\beta^{-1} \right] \right).
\end{align}
If $\widehat{H}_A=c \mathbbm{1}_A$, $ \tr \left[ (\gamma^\beta_A)^{-1} \right] = \tr (|A| \mathbbm{1}_A) = |A|^2$ and $F^{\beta}_{\infty}[\mc{U}]=2\beta^{-1}\ln|A|$.
\end{proof}
\begin{lemma}[\cite{BGD25}]\label{lemma:free-max-energy}
Given a unitary channel $\mc{U}_{A'\to A}$, the minimum value of its max-free energy $F^\beta_\infty[\mc{U}]$ is achieved for the output with trivial Hamiltonian, $\widehat{H}_A=c\mathbbm{1}_A$. In particular,
    \begin{align}
\min_{\widehat{H}_A}F^\beta_\infty[\mc{U}]=2\beta^{-1}\ln|A|.
    \end{align}
\end{lemma}
\begin{proof}
From Lemma~\ref{prop:unitary_free_energy} we have 
\begin{align}
\beta F_\infty^\beta[\mc{U}]&=\ln\tr(\gamma_\beta^{-1})\\
&=\ln \tr {\operatorname{e}}^{\beta \widehat{H}_A}+\ln\tr {\operatorname{e}}^{-\beta \widehat{H}_A}.
\end{align}
We analyze the vanishing conditions of the directional derivative with respect to $\widehat{H}_A$, that is $\nabla_{\widehat{H}_A}F_\infty^\beta[\mc{U}]=0$, which implies that
\begin{align}
 \frac{{\operatorname{e}}^{\beta\widehat{H}_A}}{\tr({\operatorname{e}}^{\beta \widehat{H}_A})}-\frac{{\operatorname{e}}^{-\beta\widehat{H}_A}}{\tr({\operatorname{e}}^{-\beta \widehat{H}_A})}=0.
\end{align}
For all eigenvalues $E_i$ of $\widehat{H}_A$, we have
\begin{align}
    \frac{{\operatorname{e}}^{\beta E_i}}{\tr({\operatorname{e}}^{\beta \widehat{H}_A})}=\frac{{\operatorname{e}}^{-\beta E_i}}{\tr({\operatorname{e}}^{-\beta \widehat{H}_A})}
\qquad    \text{or},~
{\operatorname{e}}^{2\beta E_i}=\frac{\tr({\operatorname{e}}^{\beta \widehat{H}_A})}{\tr({\operatorname{e}}^{-\beta \widehat{H}_A})},~\forall i.
\end{align}
Since the right-hand side is independent of $i$, we conclude that all the eigenvalues of $\widehat{H}_A$ must be the same, which corresponds to the Hamiltonian of the form $\widehat{H}_A=c\mathbbm{1}_A$. For such Hamiltonian, we have $F_\infty^\beta[\mc{U}]=2\beta^{-1}\ln|A|$. To see that this value is the global minimum for $F_{\infty}^\beta[\mc{U}]$, note that
\begin{align}
    {\rm e}^{\beta F_\infty^\beta[\mc{U}]}&=\tr {\operatorname{e}}^{\beta \widehat{H}_A}\tr\operatorname{e}^{-\beta \widehat{H}_A}\\
    &\ge (\tr {\operatorname{e}}^{\frac{1}{2}\beta \widehat{H}_A}{\operatorname{e}}^{-\frac{1}{2}\beta \widehat{H}_A})^2\\
    &=(\tr\mathbbm{1}_A)^2=|A|^2.
\end{align}
We have used the Cauchy-Schwarz inequality for Hilbert-Schmidt inner product: $(\tr(A^\dagger B))^2\le\tr(A^\dagger A)\tr(B^\dagger B)$. We conclude that $F_\infty^\beta[\mc{U}]\ge 2\beta^{-1}\ln|A|$ and the output Hamiltonian $\widehat{H}_A=c\mathbbm{1}_A$ gives the minimum free energy for unitary channels.
\end{proof}
The corollary given below follows directly from the Eq.~\eqref{eq:chan-work}.
\begin{corollary}\label{cor:lower-bound}
    The free energy ${F}^{\beta}[\n]$ of a quantum channel $\n_{A'\to A}$ is lower bounded as
    \begin{equation}
        F^{\beta}[\n]\geq \beta^{-1}I(R;A)_{\Phi^{\n}}+F^\beta(\n(\pi_{A'})),
    \end{equation}
    and upper bounded as
    \begin{equation}
        F^{\beta}[\n]\leq \beta^{-1}I[\n]+\sup_{\psi\in\St(A')}F^{\beta}(\mc{N}(\psi)).
    \end{equation}
    The free energy of a unitary channel $\mc{U}_{A'\to A}$ is lower bounded as
    \begin{equation}
        F^\beta[\mc{U}]\geq \beta^{-1}\left(\ln|A|+\ln Z^{\beta}_A\right)+\frac{1}{|A|}\tr[\widehat{H}_A],
    \end{equation}
    where $Z^{\beta}_A:= \tr[\exp(-\beta\widehat{H}_A)]$.
\end{corollary}
We will now show that given a unitary channel $\mc{U}_{A'\to A}$, the resource $(\mc{U},\mc{T}^\beta)$ is minimum when the output Hamiltonian is of the form $\widehat{H}_A=c\mathbbm{1}_A$. 
\begin{proposition}[Free energy of the golden unit]\label{prop:min-free-en}
Given a unitary channel $\mc{U}_{A'\to A}$, the minimum free energy of the channel is achieved for the output system Hamiltonian $\widehat{H}_A=c\mathbbm{1}_A$, In particular,
    \begin{align}
\min_{\widehat{H}_A}F^\beta_\alpha[\mc{U}]=2\beta^{-1}\ln|A|,~\text{where}~\alpha\in \{1,\infty\}.
    \end{align}
\end{proposition}
\begin{proof}
From Corollary~\ref{cor:lower-bound}, we have $F^\beta[\mc{U}]\ge f(\widehat{H}_A)$ where 
\begin{align}
f(\widehat{H}_A)=\beta^{-1}(\ln|A|+\ln Z_A^\beta+\frac{\beta}{|A|}\tr(\widehat{H}_A)).
\end{align}
To find the global minima of the above operator function, we analyze the operator derivative $\nabla_{\widehat{H}_A}f(\widehat{H}_A)$. Using arguments similar to Lemma~\ref{lemma:free-max-energy}, we can show that
\begin{align}
\min_{\widehat{H}_A}f(\widehat{H}_A)=f(c\mathbbm{1}_A)=2\beta^{-1}\ln|A|.
\end{align}
Combining the above result with that of Lemma~\ref{lemma:free-max-energy} we have the
\begin{align}
    2\beta^{-1}\ln|A|\le&\min_{\widehat{H}_A} F^\beta[\mc{U}]\le \min_{\widehat{H}_A}F_\infty^\beta[\mc{U}]=  2\beta^{-1}\ln|A|\\
  \text{This implies,}\hspace{1cm}  \min_{\widehat{H}_A} F^\beta[\mc{U}]= &\min_{\widehat{H}_A}F_\alpha^\beta[\mc{U}]=  2\beta^{-1}\ln|A|.
\end{align}
\end{proof}
With freedom of choice for golden units and the fact that GPSCs are free operations in our athermality framework, we mark our golden unit as  the identity resource channel $(\id_{A'\to A},\mathcal{R}^{\pi})$, where $\mc{R}^{\pi}_{A'\to A}$ is the absolutely thermal channel for fully degenerate output Hamiltonian ($\widehat{H}_A\propto \mathbbm{1}_A$). We denote $(\id_{A'\to A},\mc{R}^\pi_{A'\to A})$ as $(\id_m,\mc{R}^\pi)$ for $|A'|=|A|=m$.

 It is evident that $(\id_{A'\to A},\mathcal{R}^{\pi}_{A'\to A})$ is more resourceful than $(\id_{B'\to B},\mathcal{R}^{\pi}_{B'\to B})$ if $|A|>|B|$~\cite{GW21,DS25,SPSD25}, therefore, we will have to be considerate of the dimension of $\id$ in the golden unit. The identity channel $\id_m$ (or any equivalent unitary channel $\mc{U}_m$) allows for the perfect distribution of $m\times m$-dimensional quantum states between two parties~\cite{Sch96,ADHW09}, equivalent to the distribution of a maximally entangled state $\Phi$ of Schmidt rank $m$~\cite{DW05,ADHW09,KDWW19}.

 \subsection{Distillation and conversion cost of dynamical athermality}\label{subsec:distillation}
The uniformly mixing channel $\mathcal{R}^\pi_{A'\to A}$, $|A'|=|A|=m$, can be obtained by the action of unitary channels $\mc{U}(\cdot)=U(\cdot){U}^\dag$ uniformly picked at random with respect to Haar measure,
\begin{equation}
    \mathcal{R}^\pi(\cdot)=\int_{U\in\mathbbm{U}(m)}{\operatorname{d}\!\mu_U}U(\cdot)U^\dag,
\end{equation}
where $\mathbbm{U}(m)$ is the set of all unitary operators $U_{A'\to A}$. We can also obtain $\mathcal{R}^\pi$ by taking a uniform mixture of Weyl unitary channels $\mathcal{W}^i(\cdot)=W^i(\cdot){W^i}^\dagger$,
\begin{align}
    \mathcal{R}^\pi(\cdot) &=\frac{1}{m^2}\sum_{i=0}^{m^2-1} {W}^i(\cdot){W^i}^\dag,
\end{align}
where $\{W^i\}_{i=0}^{m^2-1}$ is the set (group) of Weyl unitaries; $\{W^i\}_{i=0}^{m^2-1}$ forms a complete orthonormal basis for the space of all linear operators acting on $m$-dimensional Hilbert space. Let $\mc{W}^0=\mathrm{\id}_m$ and $\id_m^{\perp}:=\sum_{i=1}^{m^2-1}\mathcal{W}^i$, then
\begin{align}
 \mc{R}^\pi  & =\frac{1}{m^2}\id_m+\frac{1}{m^2}\sum_{i=1}^{m^2-1}\mathcal{W}^i=\frac{1}{m^2}(\id_m+\id_m^{\perp}).
\end{align}
This implies~(cf.~\cite[Proposition 11]{DKSW18})
\begin{equation}
    \frac{1}{2}\norm{\id_m-\mc{R}^\pi}_{\diamond}=\frac{1}{2}\norm{\mc{U}-\mc{R}^\pi}_{\diamond}=1-\frac{1}{m^2}.
\end{equation}

\textit{Channel conversion distance}.-- For a given resource channel $(\n,\mathcal{T}^{\beta}_1)$ being transformed to another resource channel $(\mc{M},\mathcal{T}^{\beta}_2)$, the conversion distance under GPSCs are defined as
    \begin{align}
      &  d_{\mathrm{GP}}((\n,\mathcal{T}^{\beta}_1)\rightarrow(\mc{M},\mathcal{T}^{\beta}_2))\nonumber=\min_{\Theta}\left\{P[\mc{M},\Theta(\n)]~:~\Theta(\mathcal{T}^{\beta}_1)=\mathcal{T}^{\beta}_2\right\}.
    \end{align}
Here $P[\mc{M},\mc{K}]$ is the purified channel distance between the channels $\mc{M}$ and $\mc{K}$ given by
\begin{align}
    P[\mc{M},\mc{K}]:=\sup_{\psi\in \St(RA')} P(\mc{M}(\psi_{RA'}),\mc{K}(\psi_{RA'}))
\end{align}
where $P(·, ·)$ is the purified distance, and it suffices to take $\psi_{RA'}$ to be pure.
\begin{definition}[One-shot athermality distillation and formation]
For an error $\varepsilon\in[0,1]$, the single-shot athermality distillation of a resource channel $(\n,\mc{T}^\beta)$ under GPSCs is defined as
\begin{align}
        \mathrm{Dist}&^\varepsilon (\n,\mathcal{T}^\beta)\nonumber:=\sup_{m}\left\{\ln m:~d_{\mathrm{GP}}((\n,\mathcal{T}^\beta)\rightarrow(\id_m,\mathcal{R}^\pi))\le\varepsilon\right\}.
    \end{align}
For an error $\varepsilon\in[0,1]$, the single-shot athermality formation (cost) of a resource channel $(\n,\mathcal{T}^\beta)$ is defined as 
    \begin{align}
        \mathrm{Cost}&^\varepsilon (\n,\mathcal{T}^\beta)\nonumber:=\inf_{m}\left\{\ln m:~    d_{\mathrm{GP}}((\id_m,\mathcal{R}^\pi)\rightarrow(\n,\mathcal{T}^\beta))\le\varepsilon\right\}.
    \end{align}
\end{definition}

The single-shot athermality distillation $\mathrm{Dist}^\varepsilon (\n,\mathcal{T}^\beta)$ provides the largest dimensional golden unit $(\id_m,\mc{R}^{\pi})$ that can be distilled from the resource channel $(\n,\mathcal{T}^\beta)$ under GPSCs, up to allowed error $\varepsilon$. We could possibly also interpret $(\ln 2)^{-1}\mathrm{Dist}^\varepsilon (\n,\mathcal{T}^\beta)$ as the maximum number of elementary golden units $(\id_2,\mc{R}^{\pi})$ distillable from the resource channel $(\n,\mathcal{T}^\beta)$ under GPSCs. The single-shot athermality formation $\mathrm{Cost}^\varepsilon (\n,\mathcal{T}^\beta)$ provides the least dimensional golden unit $(\id_m,\mc{R}^{\pi})$ needed to form the resource channel $(\n,\mathcal{T}^\beta)$ under GPSCs, up to allowed error $\varepsilon$. Therefore, we can interpret $(\ln 2)^{-1}\mathrm{Cost}^\varepsilon (\n,\mathcal{T}^\beta)$ as the minimum number of elementary golden units $(\id_2,\mc{R}^{\pi})$ spent to form the resource channel $(\n,\mathcal{T}^\beta)$ under GPSCs.

\begin{theorem}[Athermality distillation and formation~\cite{BGD25}]\label{thm:dist_cost}
For any error $\varepsilon\in[0,1]$ and a given resource channel $(\n,\mathcal{T}^\beta)$, the single-shot athermality distillation and formation are proportional to the $\varepsilon$-hypothesis-testing free energy and the $\varepsilon$-max-free energy of the channel $\n$, respectively,
\begin{align}
    \mathrm{Dist}^\varepsilon (\n,\mathcal{T}^\beta)&=\frac{1}{2}D_{H}^{\varepsilon^2}[\n\Vert\mathcal{T}^\beta],\\
     \mathrm{Cost}^\varepsilon (\n,\mathcal{T}^\beta)&=\frac{1}{2}D_{\infty}^\varepsilon[\n\Vert\mathcal{T}^\beta],
\end{align}
$F^{\beta,\varepsilon}_H[\n]=\frac{2}{\beta}\mathrm{Dist}^{\sqrt{\varepsilon}} (\n,\mathcal{T}^\beta)$ and $F^{\beta,\varepsilon}_{\infty}[\n]=\frac{2}{\beta}\mathrm{Cost}^\varepsilon (\n,\mathcal{T}^\beta)$.
\end{theorem}
\begin{proof}
Any process describing the distillation of a golden unit can be written as a measure and prepare process where an object is processed with a free operation and then a free measurement is performed; the optimal distillation rate is when the processing with free operation and measurement is optimized for maximal distillation of the golden unit. Such a process for channel distillation of the golden unit $(\id_m,\mathcal{R}^\pi)$ from the resource object $(\mathcal{N}_{A'\to A},
\mathcal{T}^{\beta}_{A'\to A})$ can be described through a Gibbs preserving superchannel (GPSC) $\Theta_{\psi}^\Lambda$, defined through a pure state $\psi_{RA'}$ and an effect operator (element of a POVM) $0\le \Lambda\le \mathbbm{1}_{RA}$, which can be written as the following preparation,
\begin{align}
    \Theta^\Lambda_{\psi}(\n)=\tr(\n(\psi)\Lambda)\id+\tr(\n(\psi)(\mathbbm{1}_A-\Lambda))\frac{1}{m^2-1}\id^{\perp}.
\end{align}
The Gibbs preserving constraint on the superchannel, i.e. $\Theta_\psi^\Lambda(\mathcal{T}^\beta)=\mathcal{R}^\pi$, can be written as $\tr(\mathcal{T}^\beta(\psi)\Lambda)=\frac{1}{m^2}$, and the conversion distance $d_{\Theta_\psi^\Lambda}((\n,\mathcal{T}^\beta)\rightarrow(\id,\mathcal{R}^\pi))$ is given by $\sqrt{1-\tr(\n(\psi)\Lambda)}$. Recalling the definition of the $\varepsilon$-single-shot distillation, we have the following:
\begin{align}
        \mathrm{Dist}^\varepsilon (\n,\mathcal{T}^\beta)&=\sup_{m,\Lambda,\psi}\{\ln m:~  \sqrt{1-\tr(\n(\psi)\Lambda)}\le\varepsilon,~\tr(\mathcal{T}^\beta(\psi)\Lambda)=\frac{1}{m^2}\}\\
        &=\sup_{\psi,\Lambda}\left\{-\frac{1}{2}\ln\tr(\mathcal{T}^\beta(\psi)\Lambda) :~  1-\tr(\n(\psi)\Lambda)\le\varepsilon^2\right\}\\
        &=\sup_{\psi}\left[-\inf_{\Lambda}\left\{\frac{1}{2}\ln\tr(\mathcal{T}^\beta(\psi)\Lambda) :~  1-\tr(\n(\psi)\Lambda)\le\varepsilon^2\right\}\right]\\
&=\sup_{\psi}\frac{1}{2}D_H^{\varepsilon^2}(\n(\psi)\Vert\mathcal{T}^\beta(\psi))\\
         &=\frac{1}{2}D_H^{\varepsilon^2}[\n\Vert \mathcal{T}^\beta].
    \end{align}

Now, to derive the $\varepsilon$-single-shot cost, we first calculate the conversion distance: 
 \begin{align}
      d_{\mathrm{GP}}((\id_m\mathcal{R}^\pi) \rightarrow(\n,\mathcal{T}^{\beta}))&=\min_{\Theta}\left\{P[\mc{N},\Theta(\id_m)]~:~\Theta(\mathcal{R}^{\pi})=\mathcal{T}^{\beta}\right\}\\
         &=\min_{\Theta}\left\{P[\mc{N},\Theta(\id_m)]~:\mathcal{T}^{\beta}= \frac{1}{m^2}\Theta(\id_m)+\frac{1}{m^2}\Theta(\id_m^\perp)\right\}\\
         &=\min_{\mc{E}}\left\{P[\mc{N},\mc{E}]~:~m^2\mathcal{T}^{\beta}\ge \mc{E}\right\}\\
         &=\min_{\mc{E}}\left\{P[\mc{N},\mc{E}]~:~2\ln m\ge D_{\infty} [\mc{E}\Vert\mc{T}^\beta]\right\},
    \end{align}
where we have replaced $\mc{E}=\Theta(\id_m)$. Note that $\mc{E}$ can be any channel such that $m^2\mc{T}^\beta\ge \mc{E}$, and is not constrained by the Gibbs preserving property of $\Theta$. This is because the quantity that we are optimizing is a function of the variable $\Theta(\id_m)$, and the range of this variable is fully captured by the constraint $m^2\mc{T}^\beta\ge \Theta(\id_m)$. The $\varepsilon$-single-shot cost can be written as
   \begin{align}
        \mathrm{Cost}^\varepsilon (\n,\mathcal{T}^\beta)
        &=\inf_{m}\left\{\ln m:~d_{\mathrm{GP}}((\id_m\mathcal{R}^\pi) \rightarrow(\n,\mathcal{T}^{\beta})) \le\varepsilon\right\}\\
        &=\inf_{m}\{\ln m:~ P[\mc{N},\mc{E}]\le\varepsilon,~2\ln m\ge D_{\infty} [\mc{E}\Vert\mc{T}^\beta]\}\\
        &=\inf_{\mc{E}}\left\{\frac{1}{2}D_{\infty}[\mathcal{E}\Vert\mathcal{T}^\beta]:~P[\mc{N},\mc{E}] \le\varepsilon\right\}\\
         &=\frac{1}{2}D^{\varepsilon}_{\infty}[\n\Vert\mathcal{T}^\beta].
    \end{align}
\end{proof}
The one-shot athermality distillation and formation of a resource channel $(\n,\mc{T}^\beta)$ are associated with how well the channel $\n$ can be discriminated from $\mc{T}^\beta$ in an asymmetric channel discrimination task~\cite{CMW16,DW19b,WBHK20,BKSD23}, where $\mc{N}$ is null hypothesis and $\mc{T}^\beta$ is alternate hypothesis. 

Based on the assistance of parallel strategies for the channel discrimination tasks (see~\cite{CMW16,FFRS20,BKSD23} for formal discussion), for an error $\varepsilon\in(0,1)$, we define the adaptive (${\rm ad}$) and nonadaptive asymptotic athermality distillation and formation rates of a resource channel $(\n,\mc{T}^\beta)$ as
\begin{align}
    \mathscr{C}_{\mathrm{distill}}^{\varepsilon}[\mathcal{N}] 
   &:= \lim_{n \to \infty} \frac{1}{n} \, \mathrm{Distill}^{\varepsilon}_{} 
   \left( \mathcal{N}^{\otimes n}, \left( \mathcal{T}^\beta \right)^{\otimes n} \right),\\
    \mathscr{C}^{\varepsilon}_{\mathrm{cost}}[\mathcal{N}] 
   &:= \lim_{n \to \infty} \frac{1}{n} \, \mathrm{Cost}^{\varepsilon}_{} 
   \left( \mathcal{N}^{\otimes n}, \left( \mathcal{T}^\beta \right)^{\otimes n} \right).
\end{align}
It directly follows from the definitions that $\mathscr{C}_{\mathrm{distill}}^{\varepsilon}[\mathcal{N}]\leq \mathscr{C}^{\varepsilon}_{\mathrm{cost}}[\mathcal{N}]$. We prove that the asymptotic distillation and formation rates are equal under nonadaptive strategies. 

\begin{theorem}[Asymptotic reversibility]\label{thm:rev}
The resource theory of athermality of quantum channels with GPSCs as free operations is asymptotically reversible under parallel uses of channels, i.e., for any $\varepsilon\in(0,1)$ and an arbitrary (square) quantum resource channel $(\n,\mc{T}^\beta)$,
  \begin{align}
       \mathscr{C}_{\mathrm{distill}}^{\sqrt{\varepsilon}}[\mathcal{N}]=\mathscr{C}_{\mathrm{cost}}^{\varepsilon}[\mathcal{N}]&=\frac{1}{2}D[\n\Vert\mc{T}^\beta]\nonumber\\
       &=\frac{\beta}{2} F^\beta[\n].
  \end{align} 
\end{theorem}
\begin{proof}
    For any $\varepsilon\in(0,1)$, we have from~\cite[Theorem 1]{CMW16},
    \begin{equation}
        \mathscr{C}_{\mathrm{distill}}^{\sqrt{\varepsilon}}[\mathcal{N}]=\frac{1}{2} D[\n\Vert\mc{T}^\beta],
    \end{equation}
  and from~\cite[Theorem 4.1]{FGR25},
    \begin{equation}
        \mathscr{C}_{\mathrm{cost}}^{\varepsilon}[\mathcal{N}]=\frac{1}{2}D[\n\Vert\mc{T}^\beta].
    \end{equation}
    We conclude the proof by recalling that $F^\beta[\n]=\beta^{-1}D[\n\|\mc{T}^\beta]$.
\end{proof}
It follows from \cite[Theorem 2]{CMW16} that for any athermality distillation scheme which seeks to distill at a rate strictly higher than $\frac{1}{2}D[\n\|\mc{T}^\beta]$, the probability of successful distillation decays to zero exponentially fast with the number of channel uses of  $\n$. That is, $\frac{1}{2}D[\n\|\mc{T}^\beta]$ also serves the strong converse bound on the asymptotic athermality distillation rate of the channel $\n$, irrespective of whether the strategy used is adaptive or nonadaptive.

\section{Informational aspects of free energy}\label{sec:aspects}
We discuss connection of the dynamical free energy with the channel capacity and the free energy of the channel output. We observe connections of the resource theory of athermality with the information processing tasks of private randomness and purity distillation.

\textit{Private randomness distillation}.---  The private randomness capacity of a channel $\n_{A'\to A}$ is the maximum rate, in an asymptotic setting of i.i.d. uses of the channel, at which the receiver accessing output $A$ can extract private randomness, against an eavesdropper accessing extension $E$ of an isometric channel extension $\mc{V}^{\n}_{A'\to AE}$ of the channel $\n$, when the sender sends states through the channel~\cite{YHW19}. The private randomness capacity is given by $P[\n]=D[\n\|\mc{R}^\pi]=\lim_{\beta\to 0^+}D[\n\|\mc{T}^\beta]$. Therefore, the private randomness capacity of a square quantum channel is twice its asymptotic thermodynamic distillation or formation rate under the action of GPSCs in the limit inverse temperature tends to vanish.

\textit{Dynamical resource theory of purity}.---  By considering $\beta\to 0^+$ or the Hamiltonian of the channel output to be proportional to $\mathbbm{1}$ in our dynamical resource theory of athermality, we obtain the dynamical resource theory of purity~\cite{YHW19,LY20,RT21,YZGZ20} where unitary channels are golden units and the uniformly mixing channels are free objects. When $\widehat{H}_A\propto \mathbbm{1}_A$, then $\mc{T}^\beta_{A'\to A}=\mc{R}^{\pi}_{A'\to A}$ as $\gamma^\beta_A=\pi_A$ due to $\widehat{H}_A$ being fully degenerate Hamiltonian, i.e., all energy eigenstates have the same eigenvalue. The asymptotic purity distillation and formation rate of a square quantum channel $\mc{N}_{A'\to A}$ is equal to $\frac{1}{2}D[\n\|\mc{R}^\pi]=\frac{1}{2}\lim_{\beta\to 0^+}D[\n\|\mc{T}^\beta]$~\cite{YHW19,GW21,YZGZ20}. 

\textit{Entropy and thermal entropy}.--- The entropy $S[\n]$ of a quantum channel $\n_{A'\to A}$ and its private randomness capacity $P[\n]$ satisfy the following duality or trade-off relation~\cite{DGP24,DS25}:
\begin{equation}
    S[\n]+P[\n]=\ln|A|.
\end{equation}
We have a similar dual relationship between the thermal entropy $S^{\beta}[\n]:=\ln Z^{\beta}_A-D[\n\|\mathcal{T}^{\beta}]$~\cite{SPSD25} and the resource-theoretic free energy $F^{\beta}[\n]$ of a quantum channel $\n_{A'\to A}$:
\begin{equation}
    S^{\beta}[\n]+\beta F^{\beta}[\n]=\ln Z^{\beta}_A,
\end{equation}
where $Z^{\beta}_A=\tr[\exp(-\beta\widehat{H}_A)]$. A similar relation exists between the min-channel entropy and the min-free energy of a channel. In particular, if the output Hamiltonian is trivial  $\widehat{H}_{A}=0$ and $\widehat{H}_{RA}=\widehat{H}_R\otimes\mathbbm{1}_A$, we have 
\begin{equation}
S_{\infty}[\n]+\beta F^{\beta}_{\infty}[\n]=\ln|A|.
\end{equation}
The min-entropy $S_{\infty}[\n]$ of a quantum channel $\n_{A'\to A}$ is associated with the decoupling ability of the channel~\cite[Theorem 1 \& Proposition 1]{BGC+25} and the environment-assisted erasure cost $W[A|E]_{\n}$ of the output $A$ of the channel when access to the environment $E$ is provided to the erasure~\cite[Theorem 3 \& Proposition 2]{BGC+25}. These operational interpretations of the min-entropy of a quantum channel $\n_{A'\to A}$ directly provide operational interpretations for the max-free energy $F^\beta_{\infty}[\n]$ of the channel when $\widehat{H}_A=0$. 
\section{Thermodynamic free energy and work extraction}\label{sec:energy}
For a quantum state $\rho_A$, the entropy $S(\rho)$ is $S(\rho):=S(A)_{\rho}:=-D(\rho_A\Vert\mathbbm{1}_A)$ and the energy $E(\rho)$ is $E(\rho)=\langle \widehat{H}_A\rangle_{\rho}=\tr[\widehat{H}_A\rho_A]$.  The entropy $S[\n]$ of a quantum channel $\n_{A'\to A}$ is defined as~\cite{DJKR06,Yua19,GW21,SPSD25} 
\begin{align}
    S[\n]&:=-D[\n\Vert\mc{R}^\mathbbm{1}]\nonumber\\
    &= \inf_{\psi\in\St(RA')}[S(RA)_{\mc{N}(\psi)}-S(R)_{\psi}],
\end{align}
where it suffices to optimize only over pure states $\psi_{RA'}$.

\begin{definition}[Thermodynamic free energy]
    The thermodynamic (non-equilibrium) free energy ${F}^{\beta}_{\rm T}[\n]$ of an arbitrary quantum channel $\n_{A'\to A}$ is defined as
    \begin{equation}
       {F}^\beta_{\rm T}[\n]:=\beta^{-1}{D}[\n\|\widehat{\mc{T}}^\beta],
    \end{equation}
    where $\widehat{\mc{T}}^\beta=\mc{R}^{\widehat{\gamma}^\beta}$, i.e., $\widehat{\mc{T}}^\beta(\rho_{A'})=\tr[\rho_{A'}]\widehat{\gamma}^\beta_A$ is a replacer map that is completely positive but not trace-preserving.
\end{definition}

The function ${D}[\n\|\widehat{\mc{T}}^\beta]=-{S}^{\beta}[\n]$, where ${S}^\beta[\n]$ is the thermal entropy of the channel $\n$~\cite{SPSD25}. We see that ${F}^{\beta}_{\rm T}[\n]=\beta^{-1}{S}^{\beta}[\n]$. 

Based on the ``channelized" entropy and free energy definitions, we define an energy function called the thermodynamic energy $E[\n]$ of a quantum channel $\n$~\cite{BGD25,DS25}.
\begin{definition}
    The thermodynamic energy $E[\n]$ of a quantum channel $\n$ is defined as 
    \begin{equation}
    E[\n]:=\sup_{\psi\in\St(RA')}[E(\mc{N}(\psi_{RA'}))-E(\psi_R)],
\end{equation}
where it suffices to optimize only over pure states $\psi_{RA'}$ and $|R|=|A'|$.
\end{definition}

\begin{proposition}
    The thermodynamic energy $E[\n]$ of a quantum channel $\n_{A'\to A}$, with the reference-output Hamiltonian as $\widehat{H}_{RA}=\mathbbm{1}_R\otimes\widehat{H}_A+\widehat{H}_R\otimes\mathbbm{1}_A+\widehat{H}^{\rm int}_{RA}$, is
    \begin{equation}    E[\mc{N}]=\sup_{\psi\in\St(RA')}\left[\langle\widehat{H}_A\rangle_{\n(\psi_{A'})}+\langle\widehat{H}^{\rm int}_{RA}\rangle_{\n(\psi_{RA'})}\right],
    \end{equation}
    where it suffices to optimize over pure states $\psi_{RA'}$.
\end{proposition}
\begin{proof}
    The proof follows from the definition of $E[\n]$ and the fact that, for state $\n_{A'\to A}(\psi_{RA'})$ and Hamiltonian $\widehat{H}_{RA}$, we have
    \begin{align}
     E(\n(\psi_{RA'}))-E(\psi_R)=&\tr[\widehat{H}_{RA}\n(\psi_{RA'})]-\tr[\widehat{H}_R\psi_R]\nonumber\\
        =& \tr[(\widehat{H}_R\otimes\mathbbm{1}_A)\n(\psi_{RA'})]+\tr[(\mathbbm{1}_R\otimes\widehat{H}_A)\n(\psi_{RA'})]\nonumber\\
        &+\tr[\widehat{H}_{RA}^{\rm int}\n(\psi_{RA'})]-\tr[\widehat{H}_R\psi_R]\nonumber\\
        =& \tr[\widehat{H}_R\psi_R]+\tr[\widehat{H}_A\n(\psi_{A'})]+\tr[\widehat{H}^{\rm int}_{RA}\n(\psi_{RA'})]-\tr[\widehat{H}_R\psi_R]\nonumber\\
        =&\tr[\widehat{H}_A\n(\psi_{A'})]+\tr[\widehat{H}^{\rm int}_{RA}\n(\psi_{RA'})].
    \end{align}
\end{proof}

The following corollary directly follows from the above proposition.
\begin{corollary}\label{cor:energy}
    For an arbitrary quantum channel $\n_{A'\to A}$, when the reference-output have noninteracting Hamiltonian $\widehat{H}_{RA}=\mathbbm{1}_R\otimes\widehat{H}_A+\widehat{H}_R\otimes\mathbbm{1}_A$ ($\widehat{H}^{\rm int}_{RA}=0$), then
    \begin{equation}
        E[\n]=\sup_{\psi\in\St(A')}\langle\widehat{H}\rangle_{\n(\psi)},
    \end{equation}
    where it suffices to consider optimization over all pure states $\psi_{A'}$.
\end{corollary}

\begin{theorem}\label{thm:fets}
    The thermal free energy $F^{\beta}_{\rm T}[\n]$ of a quantum channel $\n_{A'\to A}$ is upper bounded as, when the reference $R$ is noninteracting with $A$ ($\widehat{H}_{RA}=\widehat{H}_R\otimes\mathbbm{1}_A+\mathbbm{1}_R\otimes\widehat{H}_A$),
    \begin{equation}
        F^{\beta}_{\rm T}[\n]\leq E[\n]-\beta^{-1}S[\n].
    \end{equation}
    The bound is saturated for the replacer channels $\mc{R}^\omega_{A'\to A}$, for $\omega\in\St(A)$ and $\widehat{H}^{\rm int}_{RA}=0$, $F^{\beta}_{\rm T}[\mc{R}^\omega]= E[\mc{R}^\omega]-\beta^{-1}S[\mc{R}^\omega]$. 
\end{theorem}
\begin{proof}
    It follows from the definition of the free energy $F^{\beta}[\n]$ that
    \begin{align}
      &  F^{\beta}[\n]=\sup_{\psi\in\St(RA')}\left[E(\n(\psi_{RA'}))-E(\psi_R)-\beta^{-1}(S(\n(\psi_{RA'}))-S(\psi_R))+\beta^{-1}(\ln Z^{\beta}_{RA}-\ln Z^\beta_R)\right],\nonumber
    \end{align}
    and $F^{\beta}[\n]=F^{\beta}_{\rm T}[\n]+\beta^{-1}\ln Z^\beta_A$. 
    
    If $\widehat{H}_{RA}=\widehat{H}_R\otimes\mathbbm{1}_A+\mathbbm{1}_R\otimes\widehat{H}_A$, then $\ln Z^\beta_{RA}=\ln Z^\beta_R+\ln Z^\beta_A$, and
    \begin{align}
       F^{\beta}_{\rm T}[\n]& =\sup_{\psi\in\St(RA')}\left[E(\n(\psi_{RA'}))-E(\psi_R)-\beta^{-1}(S(\n(\psi_{RA'}))-S(\psi_R))\right]\nonumber\\
        & \leq E[\n]-\beta^{-1}S[\n].
    \end{align}
    We conclude the proof by observing that for a replacer channel $\mc{R}^\omega$, $F^{\beta}_{\rm T}[\mc{R}^\omega]=F^{\beta}_{\rm T}(\omega)= E(\omega)-\beta^{-1}S(\omega)=E[\mc{R}^\omega]-\beta^{-1}S[\mc{R}^\omega]$, using the fact that $R$ and $A$ are noninteracting.
\end{proof}
\bibliography{output}

\begin{thebibliography}{67}%
\makeatletter
\providecommand \@ifxundefined [1]{%
 \@ifx{#1\undefined}
}%
\providecommand \@ifnum [1]{%
 \ifnum #1\expandafter \@firstoftwo
 \else \expandafter \@secondoftwo
 \fi
}%
\providecommand \@ifx [1]{%
 \ifx #1\expandafter \@firstoftwo
 \else \expandafter \@secondoftwo
 \fi
}%
\providecommand \natexlab [1]{#1}%
\providecommand \enquote  [1]{``#1''}%
\providecommand \bibnamefont  [1]{#1}%
\providecommand \bibfnamefont [1]{#1}%
\providecommand \citenamefont [1]{#1}%
\providecommand \href@noop [0]{\@secondoftwo}%
\providecommand \href [0]{\begingroup \@sanitize@url \@href}%
\providecommand \@href[1]{\@@startlink{#1}\@@href}%
\providecommand \@@href[1]{\endgroup#1\@@endlink}%
\providecommand \@sanitize@url [0]{\catcode `\\12\catcode `\$12\catcode `\&12\catcode `\#12\catcode `\^12\catcode `\_12\catcode `\%12\relax}%
\providecommand \@@startlink[1]{}%
\providecommand \@@endlink[0]{}%
\providecommand \url  [0]{\begingroup\@sanitize@url \@url }%
\providecommand \@url [1]{\endgroup\@href {#1}{\urlprefix }}%
\providecommand \urlprefix  [0]{URL }%
\providecommand \Eprint [0]{\href }%
\providecommand \doibase [0]{https://doi.org/}%
\providecommand \selectlanguage [0]{\@gobble}%
\providecommand \bibinfo  [0]{\@secondoftwo}%
\providecommand \bibfield  [0]{\@secondoftwo}%
\providecommand \translation [1]{[#1]}%
\providecommand \BibitemOpen [0]{}%
\providecommand \bibitemStop [0]{}%
\providecommand \bibitemNoStop [0]{.\EOS\space}%
\providecommand \EOS [0]{\spacefactor3000\relax}%
\providecommand \BibitemShut  [1]{\csname bibitem#1\endcsname}%
\let\auto@bib@innerbib\@empty
\bibitem [{\citenamefont {Sagawa}(2012)}]{Sag12}%
  \BibitemOpen
  \bibfield  {author} {\bibinfo {author} {\bibfnamefont {T.}~\bibnamefont {Sagawa}},\ }\bibfield  {title} {\bibinfo {title} {Thermodynamics of information processing in small systems},\ }\href {https://doi.org/https://doi.org/10.1143/PTP.127.1} {\bibfield  {journal} {\bibinfo  {journal} {Progress of theoretical physics}\ }\textbf {\bibinfo {volume} {127}},\ \bibinfo {pages} {1} (\bibinfo {year} {2012})}\BibitemShut {NoStop}%
\bibitem [{\citenamefont {Auff\`eves}(2022)}]{Auf22}%
  \BibitemOpen
  \bibfield  {author} {\bibinfo {author} {\bibfnamefont {A.}~\bibnamefont {Auff\`eves}},\ }\bibfield  {title} {\bibinfo {title} {Quantum technologies need a quantum energy initiative},\ }\href {https://doi.org/10.1103/PRXQuantum.3.020101} {\bibfield  {journal} {\bibinfo  {journal} {PRX Quantum}\ }\textbf {\bibinfo {volume} {3}},\ \bibinfo {pages} {020101} (\bibinfo {year} {2022})}\BibitemShut {NoStop}%
\bibitem [{\citenamefont {Potts}(2024)}]{Pot24}%
  \BibitemOpen
  \bibfield  {author} {\bibinfo {author} {\bibfnamefont {P.~P.}\ \bibnamefont {Potts}},\ }\href {https://arxiv.org/abs/2406.19206} {\bibinfo {title} {Quantum thermodynamics}} (\bibinfo {year} {2024}),\ \Eprint {https://arxiv.org/abs/2406.19206} {arXiv:2406.19206 [quant-ph]} \BibitemShut {NoStop}%
\bibitem [{\citenamefont {Badhani}\ and\ \citenamefont {Das}(2026)}]{BD26}%
  \BibitemOpen
  \bibfield  {author} {\bibinfo {author} {\bibfnamefont {H.}~\bibnamefont {Badhani}}\ and\ \bibinfo {author} {\bibfnamefont {S.}~\bibnamefont {Das}},\ }\href {https://arxiv.org/abs/2604.01217} {\bibinfo {title} {Conditional channel entropy sets fundamental limits on thermodynamic quantum information processing}} (\bibinfo {year} {2026}),\ \Eprint {https://arxiv.org/abs/2604.01217} {arXiv:2604.01217 [quant-ph]} \BibitemShut {NoStop}%
\bibitem [{\citenamefont {Gour}\ \emph {et~al.}(2015)\citenamefont {Gour}, \citenamefont {Müller}, \citenamefont {Narasimhachar}, \citenamefont {Spekkens},\ and\ \citenamefont {Yunger~Halpern}}]{GMN+15}%
  \BibitemOpen
  \bibfield  {author} {\bibinfo {author} {\bibfnamefont {G.}~\bibnamefont {Gour}}, \bibinfo {author} {\bibfnamefont {M.~P.}\ \bibnamefont {Müller}}, \bibinfo {author} {\bibfnamefont {V.}~\bibnamefont {Narasimhachar}}, \bibinfo {author} {\bibfnamefont {R.~W.}\ \bibnamefont {Spekkens}},\ and\ \bibinfo {author} {\bibfnamefont {N.}~\bibnamefont {Yunger~Halpern}},\ }\bibfield  {title} {\bibinfo {title} {The resource theory of informational nonequilibrium in thermodynamics},\ }\href {https://doi.org/10.1016/j.physrep.2015.04.003} {\bibfield  {journal} {\bibinfo  {journal} {Physics Reports}\ }\textbf {\bibinfo {volume} {583}},\ \bibinfo {pages} {1–58} (\bibinfo {year} {2015})}\BibitemShut {NoStop}%
\bibitem [{\citenamefont {Alicki}\ and\ \citenamefont {Horodecki}(2019)}]{AH19}%
  \BibitemOpen
  \bibfield  {author} {\bibinfo {author} {\bibfnamefont {R.}~\bibnamefont {Alicki}}\ and\ \bibinfo {author} {\bibfnamefont {M.}~\bibnamefont {Horodecki}},\ }\bibfield  {title} {\bibinfo {title} {Information-thermodynamics link revisited},\ }\href {https://doi.org/10.1088/1751-8121/ab076f} {\bibfield  {journal} {\bibinfo  {journal} {Journal of Physics A: Mathematical and Theoretical}\ }\textbf {\bibinfo {volume} {52}},\ \bibinfo {pages} {204001} (\bibinfo {year} {2019})}\BibitemShut {NoStop}%
\bibitem [{\citenamefont {Shiraishi}\ and\ \citenamefont {Sagawa}(2021)}]{SS21}%
  \BibitemOpen
  \bibfield  {author} {\bibinfo {author} {\bibfnamefont {N.}~\bibnamefont {Shiraishi}}\ and\ \bibinfo {author} {\bibfnamefont {T.}~\bibnamefont {Sagawa}},\ }\bibfield  {title} {\bibinfo {title} {Quantum thermodynamics of correlated-catalytic state conversion at small scale},\ }\href {https://doi.org/10.1103/PhysRevLett.126.150502} {\bibfield  {journal} {\bibinfo  {journal} {Physical Review Letters}\ }\textbf {\bibinfo {volume} {126}},\ \bibinfo {pages} {150502} (\bibinfo {year} {2021})}\BibitemShut {NoStop}%
\bibitem [{\citenamefont {Singh}\ \emph {et~al.}(2021)\citenamefont {Singh}, \citenamefont {Das},\ and\ \citenamefont {Cerf}}]{SDC21}%
  \BibitemOpen
  \bibfield  {author} {\bibinfo {author} {\bibfnamefont {U.}~\bibnamefont {Singh}}, \bibinfo {author} {\bibfnamefont {S.}~\bibnamefont {Das}},\ and\ \bibinfo {author} {\bibfnamefont {N.~J.}\ \bibnamefont {Cerf}},\ }\bibfield  {title} {\bibinfo {title} {Partial order on passive states and {H}offman majorization in quantum thermodynamics},\ }\href {https://doi.org/10.1103/PhysRevResearch.3.033091} {\bibfield  {journal} {\bibinfo  {journal} {Physical Review Research}\ }\textbf {\bibinfo {volume} {3}},\ \bibinfo {pages} {033091} (\bibinfo {year} {2021})}\BibitemShut {NoStop}%
\bibitem [{\citenamefont {Badhani}\ \emph {et~al.}(2025)\citenamefont {Badhani}, \citenamefont {S},\ and\ \citenamefont {Das}}]{BGD25}%
  \BibitemOpen
  \bibfield  {author} {\bibinfo {author} {\bibfnamefont {H.}~\bibnamefont {Badhani}}, \bibinfo {author} {\bibfnamefont {D.~G.}\ \bibnamefont {S}},\ and\ \bibinfo {author} {\bibfnamefont {S.}~\bibnamefont {Das}},\ }\href {https://arxiv.org/abs/2510.12790} {\bibinfo {title} {Thermodynamics of quantum processes: An operational framework for free energy and reversible athermality}} (\bibinfo {year} {2025}),\ \Eprint {https://arxiv.org/abs/2510.12790v2} {arXiv:2510.12790v2 [quant-ph]} \BibitemShut {NoStop}%
\bibitem [{\citenamefont {Bennett}(2003)}]{Ben03}%
  \BibitemOpen
  \bibfield  {author} {\bibinfo {author} {\bibfnamefont {C.~H.}\ \bibnamefont {Bennett}},\ }\bibfield  {title} {\bibinfo {title} {Notes on {L}andauer's principle, reversible computation, and {M}axwell's demon},\ }\href {https://doi.org/https://doi.org/10.1016/S1355-2198(03)00039-X} {\bibfield  {journal} {\bibinfo  {journal} {Studies in History and Philosophy of Science Part B: Studies in History and Philosophy of Modern Physics}\ }\textbf {\bibinfo {volume} {34}},\ \bibinfo {pages} {501} (\bibinfo {year} {2003})},\ \bibinfo {note} {{Q}uantum Information and Computation}\BibitemShut {NoStop}%
\bibitem [{\citenamefont {Gour}(2025)}]{Gou24}%
  \BibitemOpen
  \bibfield  {author} {\bibinfo {author} {\bibfnamefont {G.}~\bibnamefont {Gour}},\ }\href {https://www.cambridge.org/core/books/quantum-resource-theories/E478C372467A1184479474EDC74D7A40} {\emph {\bibinfo {title} {Quantum Resource Theories}}}\ (\bibinfo  {publisher} {Cambridge University Press},\ \bibinfo {year} {2025})\ \Eprint {https://arxiv.org/abs/arXiv:2402.05474} {arXiv:2402.05474} \BibitemShut {NoStop}%
\bibitem [{\citenamefont {Das}\ and\ \citenamefont {Sen}(2026)}]{DS25}%
  \BibitemOpen
  \bibfield  {author} {\bibinfo {author} {\bibfnamefont {S.}~\bibnamefont {Das}}\ and\ \bibinfo {author} {\bibfnamefont {U.}~\bibnamefont {Sen}},\ }\bibfield  {title} {\bibinfo {title} {Maximum entropy principle for quantum processes},\ }\href {https://doi.org/10.1088/1751-8121/ae7ac9} {\bibfield  {journal} {\bibinfo  {journal} {Journal of Physics A: Mathematical and Theoretical}\ }\textbf {\bibinfo {volume} {59}},\ \bibinfo {pages} {245309} (\bibinfo {year} {2026})},\ \bibinfo {note} {arXiv:2506.24079}\BibitemShut {NoStop}%
\bibitem [{\citenamefont {Horodecki}\ \emph {et~al.}(2025)\citenamefont {Horodecki}, \citenamefont {Winczewski}, \citenamefont {Sikorski}, \citenamefont {Mazurek}, \citenamefont {Czechlewski},\ and\ \citenamefont {Yehia}}]{HWS+25}%
  \BibitemOpen
  \bibfield  {author} {\bibinfo {author} {\bibfnamefont {K.}~\bibnamefont {Horodecki}}, \bibinfo {author} {\bibfnamefont {M.}~\bibnamefont {Winczewski}}, \bibinfo {author} {\bibfnamefont {L.}~\bibnamefont {Sikorski}}, \bibinfo {author} {\bibfnamefont {P.}~\bibnamefont {Mazurek}}, \bibinfo {author} {\bibfnamefont {M.}~\bibnamefont {Czechlewski}},\ and\ \bibinfo {author} {\bibfnamefont {R.}~\bibnamefont {Yehia}},\ }\href {https://arxiv.org/abs/2507.23108} {\bibinfo {title} {Quantification of the energy consumption of entanglement distribution}} (\bibinfo {year} {2025}),\ \Eprint {https://arxiv.org/abs/2507.23108} {arXiv:2507.23108 [quant-ph]} \BibitemShut {NoStop}%
\bibitem [{\citenamefont {Chiribella}\ \emph {et~al.}(2009)\citenamefont {Chiribella}, \citenamefont {D'Ariano},\ and\ \citenamefont {Perinotti}}]{CDP09}%
  \BibitemOpen
  \bibfield  {author} {\bibinfo {author} {\bibfnamefont {G.}~\bibnamefont {Chiribella}}, \bibinfo {author} {\bibfnamefont {G.~M.}\ \bibnamefont {D'Ariano}},\ and\ \bibinfo {author} {\bibfnamefont {P.}~\bibnamefont {Perinotti}},\ }\bibfield  {title} {\bibinfo {title} {Theoretical framework for quantum networks},\ }\href {https://doi.org/10.1103/PhysRevA.80.022339} {\bibfield  {journal} {\bibinfo  {journal} {Physical Review A}\ }\textbf {\bibinfo {volume} {80}},\ \bibinfo {pages} {022339} (\bibinfo {year} {2009})}\BibitemShut {NoStop}%
\bibitem [{\citenamefont {Das}(2019)}]{Das19}%
  \BibitemOpen
  \bibfield  {author} {\bibinfo {author} {\bibfnamefont {S.}~\bibnamefont {Das}},\ }\href {https://arxiv.org/abs/1901.05895} {\bibinfo {title} {Bipartite quantum interactions: Entangling and information processing abilities}} (\bibinfo {year} {2019}),\ \bibinfo {note} {arXiv:1901.05895}\BibitemShut {NoStop}%
\bibitem [{\citenamefont {Das}\ \emph {et~al.}(2021)\citenamefont {Das}, \citenamefont {B\"auml}, \citenamefont {Winczewski},\ and\ \citenamefont {Horodecki}}]{DBWH21}%
  \BibitemOpen
  \bibfield  {author} {\bibinfo {author} {\bibfnamefont {S.}~\bibnamefont {Das}}, \bibinfo {author} {\bibfnamefont {S.}~\bibnamefont {B\"auml}}, \bibinfo {author} {\bibfnamefont {M.}~\bibnamefont {Winczewski}},\ and\ \bibinfo {author} {\bibfnamefont {K.}~\bibnamefont {Horodecki}},\ }\bibfield  {title} {\bibinfo {title} {Universal limitations on quantum key distribution over a network},\ }\href {https://doi.org/10.1103/PhysRevX.11.041016} {\bibfield  {journal} {\bibinfo  {journal} {Physical Review X}\ }\textbf {\bibinfo {volume} {11}},\ \bibinfo {pages} {041016} (\bibinfo {year} {2021})}\BibitemShut {NoStop}%
\bibitem [{\citenamefont {Sohail}\ \emph {et~al.}(2026)\citenamefont {Sohail}, \citenamefont {Pandey}, \citenamefont {Singh},\ and\ \citenamefont {Das}}]{SPSD25}%
  \BibitemOpen
  \bibfield  {author} {\bibinfo {author} {\bibnamefont {Sohail}}, \bibinfo {author} {\bibfnamefont {V.}~\bibnamefont {Pandey}}, \bibinfo {author} {\bibfnamefont {U.}~\bibnamefont {Singh}},\ and\ \bibinfo {author} {\bibfnamefont {S.}~\bibnamefont {Das}},\ }\bibfield  {title} {\bibinfo {title} {Fundamental limitations on the recoverability of quantum processes},\ }\href {https://doi.org/10.1007/s00023-025-01590-y} {\bibfield  {journal} {\bibinfo  {journal} {Annales Henri Poincar{\'e}}\ }\textbf {\bibinfo {volume} {27}},\ \bibinfo {pages} {933} (\bibinfo {year} {2026})}\BibitemShut {NoStop}%
\bibitem [{\citenamefont {Das}\ \emph {et~al.}(2024)\citenamefont {Das}, \citenamefont {Goswami},\ and\ \citenamefont {Pandey}}]{DGP24}%
  \BibitemOpen
  \bibfield  {author} {\bibinfo {author} {\bibfnamefont {S.}~\bibnamefont {Das}}, \bibinfo {author} {\bibfnamefont {K.}~\bibnamefont {Goswami}},\ and\ \bibinfo {author} {\bibfnamefont {V.}~\bibnamefont {Pandey}},\ }\href {https://arxiv.org/abs/2410.01740} {\bibinfo {title} {Conditional entropy and information of quantum processes}} (\bibinfo {year} {2024}),\ \bibinfo {note} {arXiv:2410.01740}\BibitemShut {NoStop}%
\bibitem [{\citenamefont {Navascu\'es}\ and\ \citenamefont {Garc\'{\i}a-Pintos}(2015)}]{NG15}%
  \BibitemOpen
  \bibfield  {author} {\bibinfo {author} {\bibfnamefont {M.}~\bibnamefont {Navascu\'es}}\ and\ \bibinfo {author} {\bibfnamefont {L.~P.}\ \bibnamefont {Garc\'{\i}a-Pintos}},\ }\bibfield  {title} {\bibinfo {title} {Nonthermal quantum channels as a thermodynamical resource},\ }\href {https://doi.org/10.1103/PhysRevLett.115.010405} {\bibfield  {journal} {\bibinfo  {journal} {Physical Review Letters}\ }\textbf {\bibinfo {volume} {115}},\ \bibinfo {pages} {010405} (\bibinfo {year} {2015})}\BibitemShut {NoStop}%
\bibitem [{\citenamefont {Faist}\ \emph {et~al.}(2021)\citenamefont {Faist}, \citenamefont {Berta},\ and\ \citenamefont {Brandao}}]{FBB21}%
  \BibitemOpen
  \bibfield  {author} {\bibinfo {author} {\bibfnamefont {P.}~\bibnamefont {Faist}}, \bibinfo {author} {\bibfnamefont {M.}~\bibnamefont {Berta}},\ and\ \bibinfo {author} {\bibfnamefont {F.~G. S.~L.}\ \bibnamefont {Brandao}},\ }\bibfield  {title} {\bibinfo {title} {Thermodynamic implementations of quantum processes},\ }\href {https://doi.org/10.1007/s00220-021-04107-w} {\bibfield  {journal} {\bibinfo  {journal} {Communications in Mathematical Physics}\ }\textbf {\bibinfo {volume} {384}},\ \bibinfo {pages} {1709–1750} (\bibinfo {year} {2021})}\BibitemShut {NoStop}%
\bibitem [{\citenamefont {Badhani}\ \emph {et~al.}(2026)\citenamefont {Badhani}, \citenamefont {G~S}, \citenamefont {Choudhary}, \citenamefont {Anand},\ and\ \citenamefont {Das}}]{BGC+25}%
  \BibitemOpen
  \bibfield  {author} {\bibinfo {author} {\bibfnamefont {H.}~\bibnamefont {Badhani}}, \bibinfo {author} {\bibfnamefont {D.}~\bibnamefont {G~S}}, \bibinfo {author} {\bibfnamefont {S.}~\bibnamefont {Choudhary}}, \bibinfo {author} {\bibfnamefont {V.}~\bibnamefont {Anand}},\ and\ \bibinfo {author} {\bibfnamefont {S.}~\bibnamefont {Das}},\ }\bibfield  {title} {\bibinfo {title} {Erasure cost of a quantum process: A thermodynamic meaning of the dynamical min-entropy},\ }\href {http://iopscience.iop.org/article/10.1088/2058-9565/ae34e2} {\bibfield  {journal} {\bibinfo  {journal} {Quantum Science and Technology}\ }\textbf {\bibinfo {volume} {11}},\ \bibinfo {pages} {015038} (\bibinfo {year} {2026})},\ \bibinfo {note} {see also arXiv:2506.05307}\BibitemShut {NoStop}%
\bibitem [{\citenamefont {Watrous}(2005)}]{Wat05}%
  \BibitemOpen
  \bibfield  {author} {\bibinfo {author} {\bibfnamefont {J.}~\bibnamefont {Watrous}},\ }\bibfield  {title} {\bibinfo {title} {Notes on super-operator norms induced by schatten norms},\ }\href {https://doi.org/https://doi.org/10.26421/QIC5.1-6} {\bibfield  {journal} {\bibinfo  {journal} {Quantum Information \& Computation}\ }\textbf {\bibinfo {volume} {5}},\ \bibinfo {pages} {58} (\bibinfo {year} {2005})},\ \Eprint {https://arxiv.org/abs/quant-ph/0411077} {arXiv:quant-ph/0411077 [quant-ph]} \BibitemShut {NoStop}%
\bibitem [{\citenamefont {Berta}\ \emph {et~al.}(2011)\citenamefont {Berta}, \citenamefont {Christandl},\ and\ \citenamefont {Renner}}]{BCR11}%
  \BibitemOpen
  \bibfield  {author} {\bibinfo {author} {\bibfnamefont {M.}~\bibnamefont {Berta}}, \bibinfo {author} {\bibfnamefont {M.}~\bibnamefont {Christandl}},\ and\ \bibinfo {author} {\bibfnamefont {R.}~\bibnamefont {Renner}},\ }\bibfield  {title} {\bibinfo {title} {The quantum reverse shannon theorem based on one-shot information theory},\ }\href {https://doi.org/10.1007/s00220-011-1309-7} {\bibfield  {journal} {\bibinfo  {journal} {Communications in Mathematical Physics}\ }\textbf {\bibinfo {volume} {306}},\ \bibinfo {pages} {579–615} (\bibinfo {year} {2011})}\BibitemShut {NoStop}%
\bibitem [{\citenamefont {Das}\ and\ \citenamefont {Wilde}(2019)}]{DW19b}%
  \BibitemOpen
  \bibfield  {author} {\bibinfo {author} {\bibfnamefont {S.}~\bibnamefont {Das}}\ and\ \bibinfo {author} {\bibfnamefont {M.~M.}\ \bibnamefont {Wilde}},\ }\bibfield  {title} {\bibinfo {title} {Quantum rebound capacity},\ }\href {https://doi.org/10.1103/PhysRevA.100.030302} {\bibfield  {journal} {\bibinfo  {journal} {Physical Review A}\ }\textbf {\bibinfo {volume} {100}},\ \bibinfo {pages} {030302} (\bibinfo {year} {2019})}\BibitemShut {NoStop}%
\bibitem [{\citenamefont {Fang}\ \emph {et~al.}(2020)\citenamefont {Fang}, \citenamefont {Fawzi}, \citenamefont {Renner},\ and\ \citenamefont {Sutter}}]{FFRS20}%
  \BibitemOpen
  \bibfield  {author} {\bibinfo {author} {\bibfnamefont {K.}~\bibnamefont {Fang}}, \bibinfo {author} {\bibfnamefont {O.}~\bibnamefont {Fawzi}}, \bibinfo {author} {\bibfnamefont {R.}~\bibnamefont {Renner}},\ and\ \bibinfo {author} {\bibfnamefont {D.}~\bibnamefont {Sutter}},\ }\bibfield  {title} {\bibinfo {title} {Chain rule for the quantum relative entropy},\ }\href {https://doi.org/10.1103/PhysRevLett.124.100501} {\bibfield  {journal} {\bibinfo  {journal} {Physical Review Letters}\ }\textbf {\bibinfo {volume} {124}},\ \bibinfo {pages} {100501} (\bibinfo {year} {2020})}\BibitemShut {NoStop}%
\bibitem [{\citenamefont {Lenard}(1978)}]{Len78}%
  \BibitemOpen
  \bibfield  {author} {\bibinfo {author} {\bibfnamefont {A.}~\bibnamefont {Lenard}},\ }\bibfield  {title} {\bibinfo {title} {Thermodynamical proof of the {G}ibbs formula for elementary quantum systems},\ }\href {https://doi.org/10.1007/BF01011769} {\bibfield  {journal} {\bibinfo  {journal} {Journal of Statistical Physics}\ }\textbf {\bibinfo {volume} {19}},\ \bibinfo {pages} {575} (\bibinfo {year} {1978})}\BibitemShut {NoStop}%
\bibitem [{\citenamefont {Brand\~ao}\ \emph {et~al.}(2013)\citenamefont {Brand\~ao}, \citenamefont {Horodecki}, \citenamefont {Oppenheim}, \citenamefont {Renes},\ and\ \citenamefont {Spekkens}}]{BHO+13}%
  \BibitemOpen
  \bibfield  {author} {\bibinfo {author} {\bibfnamefont {F.~G. S.~L.}\ \bibnamefont {Brand\~ao}}, \bibinfo {author} {\bibfnamefont {M.}~\bibnamefont {Horodecki}}, \bibinfo {author} {\bibfnamefont {J.}~\bibnamefont {Oppenheim}}, \bibinfo {author} {\bibfnamefont {J.~M.}\ \bibnamefont {Renes}},\ and\ \bibinfo {author} {\bibfnamefont {R.~W.}\ \bibnamefont {Spekkens}},\ }\bibfield  {title} {\bibinfo {title} {Resource theory of quantum states out of thermal equilibrium},\ }\href {https://doi.org/10.1103/PhysRevLett.111.250404} {\bibfield  {journal} {\bibinfo  {journal} {Physical Review Letters}\ }\textbf {\bibinfo {volume} {111}},\ \bibinfo {pages} {250404} (\bibinfo {year} {2013})}\BibitemShut {NoStop}%
\bibitem [{\citenamefont {Deffner}\ and\ \citenamefont {Campbell}(2019)}]{DC19}%
  \BibitemOpen
  \bibfield  {author} {\bibinfo {author} {\bibfnamefont {S.}~\bibnamefont {Deffner}}\ and\ \bibinfo {author} {\bibfnamefont {S.}~\bibnamefont {Campbell}},\ }\href {https://doi.org/10.1088/2053-2571/ab21c6} {\emph {\bibinfo {title} {Quantum Thermodynamics}}},\ 2053-2571\ (\bibinfo  {publisher} {Morgan \& Claypool Publishers},\ \bibinfo {year} {2019})\ \Eprint {https://arxiv.org/abs/arXiv:1907.01596} {arXiv:1907.01596} \BibitemShut {NoStop}%
\bibitem [{\citenamefont {Brand{\~a}o}\ \emph {et~al.}(2015)\citenamefont {Brand{\~a}o}, \citenamefont {Horodecki}, \citenamefont {Ng}, \citenamefont {Oppenheim},\ and\ \citenamefont {Wehner}}]{BHN+15}%
  \BibitemOpen
  \bibfield  {author} {\bibinfo {author} {\bibfnamefont {F.~G. S.~L.}\ \bibnamefont {Brand{\~a}o}}, \bibinfo {author} {\bibfnamefont {M.}~\bibnamefont {Horodecki}}, \bibinfo {author} {\bibfnamefont {N.~H.~Y.}\ \bibnamefont {Ng}}, \bibinfo {author} {\bibfnamefont {J.}~\bibnamefont {Oppenheim}},\ and\ \bibinfo {author} {\bibfnamefont {S.}~\bibnamefont {Wehner}},\ }\bibfield  {title} {\bibinfo {title} {The second laws of quantum thermodynamics},\ }\href {https://doi.org/10.1073/pnas.1411728112} {\bibfield  {journal} {\bibinfo  {journal} {Proceedings of the National Academy of Sciences}\ }\textbf {\bibinfo {volume} {112}},\ \bibinfo {pages} {3275} (\bibinfo {year} {2015})}\BibitemShut {NoStop}%
\bibitem [{\citenamefont {Esposito}\ and\ \citenamefont {Van~den Broeck}(2011)}]{EV11}%
  \BibitemOpen
  \bibfield  {author} {\bibinfo {author} {\bibfnamefont {M.}~\bibnamefont {Esposito}}\ and\ \bibinfo {author} {\bibfnamefont {C.}~\bibnamefont {Van~den Broeck}},\ }\bibfield  {title} {\bibinfo {title} {Second law and {L}andauer principle far from equilibrium},\ }\href {https://doi.org/10.1209/0295-5075/95/40004} {\bibfield  {journal} {\bibinfo  {journal} {EPL (Europhysics Letters)}\ }\textbf {\bibinfo {volume} {95}},\ \bibinfo {pages} {40004} (\bibinfo {year} {2011})}\BibitemShut {NoStop}%
\bibitem [{\citenamefont {Chitambar}\ and\ \citenamefont {Gour}(2019)}]{CG19}%
  \BibitemOpen
  \bibfield  {author} {\bibinfo {author} {\bibfnamefont {E.}~\bibnamefont {Chitambar}}\ and\ \bibinfo {author} {\bibfnamefont {G.}~\bibnamefont {Gour}},\ }\bibfield  {title} {\bibinfo {title} {Quantum resource theories},\ }\href {https://doi.org/10.1103/RevModPhys.91.025001} {\bibfield  {journal} {\bibinfo  {journal} {Reviews of Modern Physics}\ }\textbf {\bibinfo {volume} {91}},\ \bibinfo {pages} {025001} (\bibinfo {year} {2019})}\BibitemShut {NoStop}%
\bibitem [{\citenamefont {Gour}(2022)}]{Gou22}%
  \BibitemOpen
  \bibfield  {author} {\bibinfo {author} {\bibfnamefont {G.}~\bibnamefont {Gour}},\ }\bibfield  {title} {\bibinfo {title} {Role of quantum coherence in thermodynamics},\ }\href {https://doi.org/10.1103/PRXQuantum.3.040323} {\bibfield  {journal} {\bibinfo  {journal} {PRX Quantum}\ }\textbf {\bibinfo {volume} {3}},\ \bibinfo {pages} {040323} (\bibinfo {year} {2022})}\BibitemShut {NoStop}%
\bibitem [{\citenamefont {Lami}(2025)}]{Lam25}%
  \BibitemOpen
  \bibfield  {author} {\bibinfo {author} {\bibfnamefont {L.}~\bibnamefont {Lami}},\ }\bibfield  {title} {\bibinfo {title} {A solution of the generalized quantum {Stein’s} lemma},\ }\href {https://doi.org/10.1109/tit.2025.3543610} {\bibfield  {journal} {\bibinfo  {journal} {IEEE Transactions on Information Theory}\ }\textbf {\bibinfo {volume} {71}},\ \bibinfo {pages} {4454–4484} (\bibinfo {year} {2025})}\BibitemShut {NoStop}%
\bibitem [{\citenamefont {Bruschi}\ \emph {et~al.}(2015)\citenamefont {Bruschi}, \citenamefont {Perarnau-Llobet}, \citenamefont {Friis}, \citenamefont {Hovhannisyan},\ and\ \citenamefont {Huber}}]{BPF+15}%
  \BibitemOpen
  \bibfield  {author} {\bibinfo {author} {\bibfnamefont {D.~E.}\ \bibnamefont {Bruschi}}, \bibinfo {author} {\bibfnamefont {M.}~\bibnamefont {Perarnau-Llobet}}, \bibinfo {author} {\bibfnamefont {N.}~\bibnamefont {Friis}}, \bibinfo {author} {\bibfnamefont {K.~V.}\ \bibnamefont {Hovhannisyan}},\ and\ \bibinfo {author} {\bibfnamefont {M.}~\bibnamefont {Huber}},\ }\bibfield  {title} {\bibinfo {title} {Thermodynamics of creating correlations: Limitations and optimal protocols},\ }\href {https://doi.org/10.1103/PhysRevE.91.032118} {\bibfield  {journal} {\bibinfo  {journal} {Physical Review E}\ }\textbf {\bibinfo {volume} {91}},\ \bibinfo {pages} {032118} (\bibinfo {year} {2015})}\BibitemShut {NoStop}%
\bibitem [{\citenamefont {Perarnau-Llobet}\ \emph {et~al.}(2015)\citenamefont {Perarnau-Llobet}, \citenamefont {Hovhannisyan}, \citenamefont {Huber}, \citenamefont {Skrzypczyk}, \citenamefont {Brunner},\ and\ \citenamefont {Ac\'{\i}n}}]{PHH+15}%
  \BibitemOpen
  \bibfield  {author} {\bibinfo {author} {\bibfnamefont {M.}~\bibnamefont {Perarnau-Llobet}}, \bibinfo {author} {\bibfnamefont {K.~V.}\ \bibnamefont {Hovhannisyan}}, \bibinfo {author} {\bibfnamefont {M.}~\bibnamefont {Huber}}, \bibinfo {author} {\bibfnamefont {P.}~\bibnamefont {Skrzypczyk}}, \bibinfo {author} {\bibfnamefont {N.}~\bibnamefont {Brunner}},\ and\ \bibinfo {author} {\bibfnamefont {A.}~\bibnamefont {Ac\'{\i}n}},\ }\bibfield  {title} {\bibinfo {title} {Extractable work from correlations},\ }\href {https://doi.org/10.1103/PhysRevX.5.041011} {\bibfield  {journal} {\bibinfo  {journal} {Physical Review X}\ }\textbf {\bibinfo {volume} {5}},\ \bibinfo {pages} {041011} (\bibinfo {year} {2015})}\BibitemShut {NoStop}%
\bibitem [{\citenamefont {Bera}\ \emph {et~al.}(2019)\citenamefont {Bera}, \citenamefont {Riera}, \citenamefont {Lewenstein}, \citenamefont {Khanian},\ and\ \citenamefont {Winter}}]{BRL+19}%
  \BibitemOpen
  \bibfield  {author} {\bibinfo {author} {\bibfnamefont {M.~N.}\ \bibnamefont {Bera}}, \bibinfo {author} {\bibfnamefont {A.}~\bibnamefont {Riera}}, \bibinfo {author} {\bibfnamefont {M.}~\bibnamefont {Lewenstein}}, \bibinfo {author} {\bibfnamefont {Z.~B.}\ \bibnamefont {Khanian}},\ and\ \bibinfo {author} {\bibfnamefont {A.}~\bibnamefont {Winter}},\ }\bibfield  {title} {\bibinfo {title} {Thermodynamics as a consequence of information conservation},\ }\href {https://doi.org/10.22331/q-2019-02-14-121} {\bibfield  {journal} {\bibinfo  {journal} {Quantum}\ }\textbf {\bibinfo {volume} {3}},\ \bibinfo {pages} {121} (\bibinfo {year} {2019})}\BibitemShut {NoStop}%
\bibitem [{\citenamefont {Devetak}\ \emph {et~al.}(2006)\citenamefont {Devetak}, \citenamefont {Junge}, \citenamefont {King},\ and\ \citenamefont {Ruskai}}]{DJKR06}%
  \BibitemOpen
  \bibfield  {author} {\bibinfo {author} {\bibfnamefont {I.}~\bibnamefont {Devetak}}, \bibinfo {author} {\bibfnamefont {M.}~\bibnamefont {Junge}}, \bibinfo {author} {\bibfnamefont {C.}~\bibnamefont {King}},\ and\ \bibinfo {author} {\bibfnamefont {M.~B.}\ \bibnamefont {Ruskai}},\ }\bibfield  {title} {\bibinfo {title} {Multiplicativity of completely bounded p-norms implies a new additivity result},\ }\href {https://doi.org/10.1007/s00220-006-0034-0} {\bibfield  {journal} {\bibinfo  {journal} {Communications in Mathematical Physics}\ }\textbf {\bibinfo {volume} {266}},\ \bibinfo {pages} {37–63} (\bibinfo {year} {2006})}\BibitemShut {NoStop}%
\bibitem [{\citenamefont {Gour}\ and\ \citenamefont {Wilde}(2021)}]{GW21}%
  \BibitemOpen
  \bibfield  {author} {\bibinfo {author} {\bibfnamefont {G.}~\bibnamefont {Gour}}\ and\ \bibinfo {author} {\bibfnamefont {M.~M.}\ \bibnamefont {Wilde}},\ }\bibfield  {title} {\bibinfo {title} {Entropy of a quantum channel},\ }\href {https://doi.org/10.1103/PhysRevResearch.3.023096} {\bibfield  {journal} {\bibinfo  {journal} {Physical Review Research}\ }\textbf {\bibinfo {volume} {3}},\ \bibinfo {pages} {023096} (\bibinfo {year} {2021})}\BibitemShut {NoStop}%
\bibitem [{Note1()}]{Note1}%
  \BibitemOpen
  \bibinfo {note} {A quantum channel is said to be square if its input and output dimensions are equal.}\BibitemShut {Stop}%
\bibitem [{\citenamefont {Cooney}\ \emph {et~al.}(2016)\citenamefont {Cooney}, \citenamefont {Mosonyi},\ and\ \citenamefont {Wilde}}]{CMW16}%
  \BibitemOpen
  \bibfield  {author} {\bibinfo {author} {\bibfnamefont {T.}~\bibnamefont {Cooney}}, \bibinfo {author} {\bibfnamefont {M.}~\bibnamefont {Mosonyi}},\ and\ \bibinfo {author} {\bibfnamefont {M.~M.}\ \bibnamefont {Wilde}},\ }\bibfield  {title} {\bibinfo {title} {Strong converse exponents for a quantum channel discrimination problem and quantum‐feedback‐assisted communication},\ }\href {https://doi.org/10.1007/s00220-016-2645-4} {\bibfield  {journal} {\bibinfo  {journal} {Communications in Mathematical Physics}\ }\textbf {\bibinfo {volume} {344}},\ \bibinfo {pages} {797} (\bibinfo {year} {2016})},\ \bibinfo {note} {arXiv:1408.3373}\BibitemShut {NoStop}%
\bibitem [{\citenamefont {Leditzky}\ \emph {et~al.}(2018)\citenamefont {Leditzky}, \citenamefont {Kaur}, \citenamefont {Datta},\ and\ \citenamefont {Wilde}}]{LKDW18}%
  \BibitemOpen
  \bibfield  {author} {\bibinfo {author} {\bibfnamefont {F.}~\bibnamefont {Leditzky}}, \bibinfo {author} {\bibfnamefont {E.}~\bibnamefont {Kaur}}, \bibinfo {author} {\bibfnamefont {N.}~\bibnamefont {Datta}},\ and\ \bibinfo {author} {\bibfnamefont {M.~M.}\ \bibnamefont {Wilde}},\ }\bibfield  {title} {\bibinfo {title} {Approaches for approximate additivity of the {H}olevo information of quantum channels},\ }\href {https://journals.aps.org/pra/abstract/10.1103/PhysRevA.97.012332} {\bibfield  {journal} {\bibinfo  {journal} {Physical Review A}\ }\textbf {\bibinfo {volume} {97}},\ \bibinfo {pages} {012332} (\bibinfo {year} {2018})}\BibitemShut {NoStop}%
\bibitem [{\citenamefont {Fawzi}\ \emph {et~al.}(2025)\citenamefont {Fawzi}, \citenamefont {Gao},\ and\ \citenamefont {Rahaman}}]{FGR25}%
  \BibitemOpen
  \bibfield  {author} {\bibinfo {author} {\bibfnamefont {O.}~\bibnamefont {Fawzi}}, \bibinfo {author} {\bibfnamefont {L.}~\bibnamefont {Gao}},\ and\ \bibinfo {author} {\bibfnamefont {M.}~\bibnamefont {Rahaman}},\ }\bibfield  {title} {\bibinfo {title} {Asymptotic equipartition theorems in von {N}eumann algebras},\ }\bibfield  {journal} {\bibinfo  {journal} {Annales Henri Poincaré}\ }\href {https://doi.org/10.1007/s00023-025-01545-3} {10.1007/s00023-025-01545-3} (\bibinfo {year} {2025})\BibitemShut {NoStop}%
\bibitem [{\citenamefont {Wang}\ and\ \citenamefont {Wilde}(2019)}]{WW19}%
  \BibitemOpen
  \bibfield  {author} {\bibinfo {author} {\bibfnamefont {X.}~\bibnamefont {Wang}}\ and\ \bibinfo {author} {\bibfnamefont {M.~M.}\ \bibnamefont {Wilde}},\ }\bibfield  {title} {\bibinfo {title} {Resource theory of asymmetric distinguishability for quantum channels},\ }\href {https://doi.org/10.1103/PhysRevResearch.1.033169} {\bibfield  {journal} {\bibinfo  {journal} {Physical Review Research}\ }\textbf {\bibinfo {volume} {1}},\ \bibinfo {pages} {033169} (\bibinfo {year} {2019})}\BibitemShut {NoStop}%
\bibitem [{\citenamefont {Yuan}\ \emph {et~al.}(2020)\citenamefont {Yuan}, \citenamefont {Zeng}, \citenamefont {Gao},\ and\ \citenamefont {Zhao}}]{YZGZ20}%
  \BibitemOpen
  \bibfield  {author} {\bibinfo {author} {\bibfnamefont {X.}~\bibnamefont {Yuan}}, \bibinfo {author} {\bibfnamefont {P.}~\bibnamefont {Zeng}}, \bibinfo {author} {\bibfnamefont {M.}~\bibnamefont {Gao}},\ and\ \bibinfo {author} {\bibfnamefont {Q.}~\bibnamefont {Zhao}},\ }\href {https://arxiv.org/abs/2012.02781} {\bibinfo {title} {One-shot dynamical resource theory}} (\bibinfo {year} {2020}),\ \Eprint {https://arxiv.org/abs/2012.02781} {arXiv:2012.02781 [quant-ph]} \BibitemShut {NoStop}%
\bibitem [{\citenamefont {Campaioli}\ \emph {et~al.}(2024)\citenamefont {Campaioli}, \citenamefont {Gherardini}, \citenamefont {Quach}, \citenamefont {Polini},\ and\ \citenamefont {Andolina}}]{CGQ+24}%
  \BibitemOpen
  \bibfield  {author} {\bibinfo {author} {\bibfnamefont {F.}~\bibnamefont {Campaioli}}, \bibinfo {author} {\bibfnamefont {S.}~\bibnamefont {Gherardini}}, \bibinfo {author} {\bibfnamefont {J.~Q.}\ \bibnamefont {Quach}}, \bibinfo {author} {\bibfnamefont {M.}~\bibnamefont {Polini}},\ and\ \bibinfo {author} {\bibfnamefont {G.~M.}\ \bibnamefont {Andolina}},\ }\bibfield  {title} {\bibinfo {title} {Colloquium: Quantum batteries},\ }\href {https://doi.org/10.1103/RevModPhys.96.031001} {\bibfield  {journal} {\bibinfo  {journal} {Reviews of Modern Physics}\ }\textbf {\bibinfo {volume} {96}},\ \bibinfo {pages} {031001} (\bibinfo {year} {2024})}\BibitemShut {NoStop}%
\bibitem [{\citenamefont {Tirone}\ \emph {et~al.}(2025)\citenamefont {Tirone}, \citenamefont {Salvia}, \citenamefont {Chessa},\ and\ \citenamefont {Giovannetti}}]{TSCG25}%
  \BibitemOpen
  \bibfield  {author} {\bibinfo {author} {\bibfnamefont {S.}~\bibnamefont {Tirone}}, \bibinfo {author} {\bibfnamefont {R.}~\bibnamefont {Salvia}}, \bibinfo {author} {\bibfnamefont {S.}~\bibnamefont {Chessa}},\ and\ \bibinfo {author} {\bibfnamefont {V.}~\bibnamefont {Giovannetti}},\ }\bibfield  {title} {\bibinfo {title} {Quantum work extraction efficiency for noisy quantum batteries: The role of coherence},\ }\href {https://doi.org/10.1103/PhysRevA.111.012204} {\bibfield  {journal} {\bibinfo  {journal} {Physical Review A}\ }\textbf {\bibinfo {volume} {111}},\ \bibinfo {pages} {012204} (\bibinfo {year} {2025})}\BibitemShut {NoStop}%
\bibitem [{\citenamefont {Chakraborty}\ \emph {et~al.}(2025)\citenamefont {Chakraborty}, \citenamefont {Das}, \citenamefont {Ghorui}, \citenamefont {Hazra},\ and\ \citenamefont {Singh}}]{CDG+25}%
  \BibitemOpen
  \bibfield  {author} {\bibinfo {author} {\bibfnamefont {S.}~\bibnamefont {Chakraborty}}, \bibinfo {author} {\bibfnamefont {S.}~\bibnamefont {Das}}, \bibinfo {author} {\bibfnamefont {A.}~\bibnamefont {Ghorui}}, \bibinfo {author} {\bibfnamefont {S.}~\bibnamefont {Hazra}},\ and\ \bibinfo {author} {\bibfnamefont {U.}~\bibnamefont {Singh}},\ }\bibfield  {title} {\bibinfo {title} {Sample complexity of black box work extraction},\ }\href {https://doi.org/10.1088/2058-9565/ae0e4d} {\bibfield  {journal} {\bibinfo  {journal} {Quantum Science and Technology}\ }\textbf {\bibinfo {volume} {10}},\ \bibinfo {pages} {045070} (\bibinfo {year} {2025})}\BibitemShut {NoStop}%
\bibitem [{\citenamefont {Yang}\ \emph {et~al.}(2019)\citenamefont {Yang}, \citenamefont {Horodecki},\ and\ \citenamefont {Winter}}]{YHW19}%
  \BibitemOpen
  \bibfield  {author} {\bibinfo {author} {\bibfnamefont {D.}~\bibnamefont {Yang}}, \bibinfo {author} {\bibfnamefont {K.}~\bibnamefont {Horodecki}},\ and\ \bibinfo {author} {\bibfnamefont {A.}~\bibnamefont {Winter}},\ }\bibfield  {title} {\bibinfo {title} {Distributed private randomness distillation},\ }\href {https://doi.org/10.1103/PhysRevLett.123.170501} {\bibfield  {journal} {\bibinfo  {journal} {Physical Review Letters}\ }\textbf {\bibinfo {volume} {123}},\ \bibinfo {pages} {170501} (\bibinfo {year} {2019})}\BibitemShut {NoStop}%
\bibitem [{\citenamefont {Shiraishi}\ and\ \citenamefont {Takagi}(2025)}]{ST25}%
  \BibitemOpen
  \bibfield  {author} {\bibinfo {author} {\bibfnamefont {N.}~\bibnamefont {Shiraishi}}\ and\ \bibinfo {author} {\bibfnamefont {R.}~\bibnamefont {Takagi}},\ }\href {https://arxiv.org/abs/2510.05642} {\bibinfo {title} {Recovery of the second law in fully quantum thermodynamics}} (\bibinfo {year} {2025}),\ \Eprint {https://arxiv.org/abs/2510.05642} {arXiv:2510.05642 [quant-ph]} \BibitemShut {NoStop}%
\bibitem [{\citenamefont {Tomamichel}(2021)}]{Tom21}%
  \BibitemOpen
  \bibfield  {author} {\bibinfo {author} {\bibfnamefont {M.}~\bibnamefont {Tomamichel}},\ }\href {https://doi.org/10.48550/arXiv.1504.00233} {\bibinfo {title} {Quantum information processing with finite resources -- mathematical foundations}} (\bibinfo {year} {2021}),\ \bibinfo {note} {arXiv:1504.00233v5}\BibitemShut {NoStop}%
\bibitem [{\citenamefont {Wilde}\ \emph {et~al.}(2014)\citenamefont {Wilde}, \citenamefont {Winter},\ and\ \citenamefont {Yang}}]{WWY14}%
  \BibitemOpen
  \bibfield  {author} {\bibinfo {author} {\bibfnamefont {M.~M.}\ \bibnamefont {Wilde}}, \bibinfo {author} {\bibfnamefont {A.}~\bibnamefont {Winter}},\ and\ \bibinfo {author} {\bibfnamefont {D.}~\bibnamefont {Yang}},\ }\bibfield  {title} {\bibinfo {title} {Strong converse for the classical capacity of entanglement-breaking and {Hadamard} channels via a sandwiched {R\'enyi} relative entropy},\ }\href {https://doi.org/10.1007/s00220-014-2057-z} {\bibfield  {journal} {\bibinfo  {journal} {Communications in Mathematical Physics}\ }\textbf {\bibinfo {volume} {331}},\ \bibinfo {pages} {593} (\bibinfo {year} {2014})},\ \bibinfo {note} {arXiv:1306.1586}\BibitemShut {NoStop}%
\bibitem [{\citenamefont {Datta}(2009)}]{Dat09}%
  \BibitemOpen
  \bibfield  {author} {\bibinfo {author} {\bibfnamefont {N.}~\bibnamefont {Datta}},\ }\bibfield  {title} {\bibinfo {title} {Min- and max-relative entropies and a new entanglement monotone},\ }\href {https://doi.org/10.1109/TIT.2009.2018325} {\bibfield  {journal} {\bibinfo  {journal} {IEEE Transactions on Information Theory}\ }\textbf {\bibinfo {volume} {55}},\ \bibinfo {pages} {2816} (\bibinfo {year} {2009})}\BibitemShut {NoStop}%
\bibitem [{\citenamefont {Wang}\ and\ \citenamefont {Renner}(2012)}]{LR12}%
  \BibitemOpen
  \bibfield  {author} {\bibinfo {author} {\bibfnamefont {L.}~\bibnamefont {Wang}}\ and\ \bibinfo {author} {\bibfnamefont {R.}~\bibnamefont {Renner}},\ }\bibfield  {title} {\bibinfo {title} {One-shot classical-quantum capacity and hypothesis testing},\ }\href {https://doi.org/10.1103/PhysRevLett.108.200501} {\bibfield  {journal} {\bibinfo  {journal} {Physical Review Letters}\ }\textbf {\bibinfo {volume} {108}},\ \bibinfo {pages} {200501} (\bibinfo {year} {2012})}\BibitemShut {NoStop}%
\bibitem [{\citenamefont {Berta}\ \emph {et~al.}(2025)\citenamefont {Berta}, \citenamefont {Lami},\ and\ \citenamefont {Tomamichel}}]{BLT25}%
  \BibitemOpen
  \bibfield  {author} {\bibinfo {author} {\bibfnamefont {M.}~\bibnamefont {Berta}}, \bibinfo {author} {\bibfnamefont {L.}~\bibnamefont {Lami}},\ and\ \bibinfo {author} {\bibfnamefont {M.}~\bibnamefont {Tomamichel}},\ }\bibfield  {title} {\bibinfo {title} {Continuity of entropies via integral representations},\ }\href {https://doi.org/10.1109/TIT.2025.3527858} {\bibfield  {journal} {\bibinfo  {journal} {IEEE Transactions on Information Theory}\ }\textbf {\bibinfo {volume} {71}},\ \bibinfo {pages} {1896} (\bibinfo {year} {2025})}\BibitemShut {NoStop}%
\bibitem [{\citenamefont {Audenaert}\ \emph {et~al.}(2025)\citenamefont {Audenaert}, \citenamefont {Bergh}, \citenamefont {Datta}, \citenamefont {Jabbour}, \citenamefont {Capel},\ and\ \citenamefont {Gondolf}}]{ABD+25}%
  \BibitemOpen
  \bibfield  {author} {\bibinfo {author} {\bibfnamefont {K.}~\bibnamefont {Audenaert}}, \bibinfo {author} {\bibfnamefont {B.}~\bibnamefont {Bergh}}, \bibinfo {author} {\bibfnamefont {N.}~\bibnamefont {Datta}}, \bibinfo {author} {\bibfnamefont {M.~G.}\ \bibnamefont {Jabbour}}, \bibinfo {author} {\bibfnamefont {{\'A}.}~\bibnamefont {Capel}},\ and\ \bibinfo {author} {\bibfnamefont {P.}~\bibnamefont {Gondolf}},\ }\bibfield  {title} {\bibinfo {title} {Continuity bounds for quantum entropies arising from a fundamental entropic inequality},\ }\href {https://doi.org/10.1109/TIT.2025.3586478} {\bibfield  {journal} {\bibinfo  {journal} {{IEEE Transactions on Information Theory}}\ }\textbf {\bibinfo {volume} {71}},\ \bibinfo {pages} {7029 } (\bibinfo {year} {2025})}\BibitemShut {NoStop}%
\bibitem [{\citenamefont {Wilde}\ \emph {et~al.}(2020)\citenamefont {Wilde}, \citenamefont {Berta}, \citenamefont {Hirche},\ and\ \citenamefont {Kaur}}]{WBHK20}%
  \BibitemOpen
  \bibfield  {author} {\bibinfo {author} {\bibfnamefont {M.~M.}\ \bibnamefont {Wilde}}, \bibinfo {author} {\bibfnamefont {M.}~\bibnamefont {Berta}}, \bibinfo {author} {\bibfnamefont {C.}~\bibnamefont {Hirche}},\ and\ \bibinfo {author} {\bibfnamefont {E.}~\bibnamefont {Kaur}},\ }\bibfield  {title} {\bibinfo {title} {Amortized channel divergence for asymptotic quantum channel discrimination},\ }\href {https://doi.org/10.1007/s11005-020-01297-7} {\bibfield  {journal} {\bibinfo  {journal} {Letters in Mathematical Physics}\ }\textbf {\bibinfo {volume} {110}},\ \bibinfo {pages} {2277–2336} (\bibinfo {year} {2020})}\BibitemShut {NoStop}%
\bibitem [{\citenamefont {Ruskai}(2002)}]{Rus02}%
  \BibitemOpen
  \bibfield  {author} {\bibinfo {author} {\bibfnamefont {M.~B.}\ \bibnamefont {Ruskai}},\ }\bibfield  {title} {\bibinfo {title} {Inequalities for quantum entropy: A review with conditions for equality},\ }\href@noop {} {\bibfield  {journal} {\bibinfo  {journal} {Journal of Mathematical Physics}\ }\textbf {\bibinfo {volume} {43}},\ \bibinfo {pages} {4358} (\bibinfo {year} {2002})}\BibitemShut {NoStop}%
\bibitem [{\citenamefont {Luo}\ \emph {et~al.}(2025)\citenamefont {Luo}, \citenamefont {Milz},\ and\ \citenamefont {Binder}}]{LMB25}%
  \BibitemOpen
  \bibfield  {author} {\bibinfo {author} {\bibfnamefont {Y.}~\bibnamefont {Luo}}, \bibinfo {author} {\bibfnamefont {S.}~\bibnamefont {Milz}},\ and\ \bibinfo {author} {\bibfnamefont {F.~C.}\ \bibnamefont {Binder}},\ }\href {https://arxiv.org/abs/2506.20428} {\bibinfo {title} {Thermodynamic criteria for signaling in quantum channels}} (\bibinfo {year} {2025}),\ \Eprint {https://arxiv.org/abs/2506.20428} {arXiv:2506.20428 [quant-ph]} \BibitemShut {NoStop}%
\bibitem [{\citenamefont {Schumacher}(1996)}]{Sch96}%
  \BibitemOpen
  \bibfield  {author} {\bibinfo {author} {\bibfnamefont {B.}~\bibnamefont {Schumacher}},\ }\bibfield  {title} {\bibinfo {title} {Sending entanglement through noisy quantum channels},\ }\href {https://doi.org/10.1103/PhysRevA.54.2614} {\bibfield  {journal} {\bibinfo  {journal} {Physical Review A}\ }\textbf {\bibinfo {volume} {54}},\ \bibinfo {pages} {2614} (\bibinfo {year} {1996})}\BibitemShut {NoStop}%
\bibitem [{\citenamefont {Abeyesinghe}\ \emph {et~al.}(2009)\citenamefont {Abeyesinghe}, \citenamefont {Devetak}, \citenamefont {Hayden},\ and\ \citenamefont {Winter}}]{ADHW09}%
  \BibitemOpen
  \bibfield  {author} {\bibinfo {author} {\bibfnamefont {A.}~\bibnamefont {Abeyesinghe}}, \bibinfo {author} {\bibfnamefont {I.}~\bibnamefont {Devetak}}, \bibinfo {author} {\bibfnamefont {P.}~\bibnamefont {Hayden}},\ and\ \bibinfo {author} {\bibfnamefont {A.}~\bibnamefont {Winter}},\ }\bibfield  {title} {\bibinfo {title} {The mother of all protocols: restructuring quantum information’s family tree},\ }\href {https://doi.org/10.1098/rspa.2009.0202} {\bibfield  {journal} {\bibinfo  {journal} {Proceedings of the Royal Society A: Mathematical, Physical and Engineering Sciences}\ }\textbf {\bibinfo {volume} {465}},\ \bibinfo {pages} {2537–2563} (\bibinfo {year} {2009})}\BibitemShut {NoStop}%
\bibitem [{\citenamefont {Devetak}\ and\ \citenamefont {Winter}(2005)}]{DW05}%
  \BibitemOpen
  \bibfield  {author} {\bibinfo {author} {\bibfnamefont {I.}~\bibnamefont {Devetak}}\ and\ \bibinfo {author} {\bibfnamefont {A.}~\bibnamefont {Winter}},\ }\bibfield  {title} {\bibinfo {title} {Distillation of secret key and entanglement from quantum states},\ }\href {https://doi.org/10.1098/rspa.2004.1372} {\bibfield  {journal} {\bibinfo  {journal} {Proceedings of the Royal Society A: Mathematical, Physical and Engineering Sciences}\ }\textbf {\bibinfo {volume} {461}},\ \bibinfo {pages} {207–235} (\bibinfo {year} {2005})}\BibitemShut {NoStop}%
\bibitem [{\citenamefont {Kaur}\ \emph {et~al.}(2019)\citenamefont {Kaur}, \citenamefont {Das}, \citenamefont {Wilde},\ and\ \citenamefont {Winter}}]{KDWW19}%
  \BibitemOpen
  \bibfield  {author} {\bibinfo {author} {\bibfnamefont {E.}~\bibnamefont {Kaur}}, \bibinfo {author} {\bibfnamefont {S.}~\bibnamefont {Das}}, \bibinfo {author} {\bibfnamefont {M.~M.}\ \bibnamefont {Wilde}},\ and\ \bibinfo {author} {\bibfnamefont {A.}~\bibnamefont {Winter}},\ }\bibfield  {title} {\bibinfo {title} {Extendibility limits the performance of quantum processors},\ }\href {https://doi.org/10.1103/PhysRevLett.123.070502} {\bibfield  {journal} {\bibinfo  {journal} {Physical Review Letters}\ }\textbf {\bibinfo {volume} {123}},\ \bibinfo {pages} {070502} (\bibinfo {year} {2019})}\BibitemShut {NoStop}%
\bibitem [{\citenamefont {Das}\ \emph {et~al.}(2018)\citenamefont {Das}, \citenamefont {Khatri}, \citenamefont {Siopsis},\ and\ \citenamefont {Wilde}}]{DKSW18}%
  \BibitemOpen
  \bibfield  {author} {\bibinfo {author} {\bibfnamefont {S.}~\bibnamefont {Das}}, \bibinfo {author} {\bibfnamefont {S.}~\bibnamefont {Khatri}}, \bibinfo {author} {\bibfnamefont {G.}~\bibnamefont {Siopsis}},\ and\ \bibinfo {author} {\bibfnamefont {M.~M.}\ \bibnamefont {Wilde}},\ }\bibfield  {title} {\bibinfo {title} {Fundamental limits on quantum dynamics based on entropy change},\ }\bibfield  {journal} {\bibinfo  {journal} {Journal of Mathematical Physics}\ }\textbf {\bibinfo {volume} {59}},\ \href {https://doi.org/10.1063/1.4997044} {10.1063/1.4997044} (\bibinfo {year} {2018})\BibitemShut {NoStop}%
\bibitem [{\citenamefont {Bergh}\ \emph {et~al.}(2023)\citenamefont {Bergh}, \citenamefont {Kochanowski}, \citenamefont {Salzmann},\ and\ \citenamefont {Datta}}]{BKSD23}%
  \BibitemOpen
  \bibfield  {author} {\bibinfo {author} {\bibfnamefont {B.}~\bibnamefont {Bergh}}, \bibinfo {author} {\bibfnamefont {J.}~\bibnamefont {Kochanowski}}, \bibinfo {author} {\bibfnamefont {R.}~\bibnamefont {Salzmann}},\ and\ \bibinfo {author} {\bibfnamefont {N.}~\bibnamefont {Datta}},\ }\href {https://arxiv.org/abs/2308.12959} {\bibinfo {title} {Infinite dimensional asymmetric quantum channel discrimination}} (\bibinfo {year} {2023}),\ \Eprint {https://arxiv.org/abs/2308.12959} {arXiv:2308.12959 [quant-ph]} \BibitemShut {NoStop}%
\bibitem [{\citenamefont {Liu}\ and\ \citenamefont {Yuan}(2020)}]{LY20}%
  \BibitemOpen
  \bibfield  {author} {\bibinfo {author} {\bibfnamefont {Y.}~\bibnamefont {Liu}}\ and\ \bibinfo {author} {\bibfnamefont {X.}~\bibnamefont {Yuan}},\ }\bibfield  {title} {\bibinfo {title} {Operational resource theory of quantum channels},\ }\bibfield  {journal} {\bibinfo  {journal} {Physical Review Research}\ }\textbf {\bibinfo {volume} {2}},\ \href {https://doi.org/10.1103/physrevresearch.2.012035} {10.1103/physrevresearch.2.012035} (\bibinfo {year} {2020})\BibitemShut {NoStop}%
\bibitem [{\citenamefont {Regula}\ and\ \citenamefont {Takagi}(2021)}]{RT21}%
  \BibitemOpen
  \bibfield  {author} {\bibinfo {author} {\bibfnamefont {B.}~\bibnamefont {Regula}}\ and\ \bibinfo {author} {\bibfnamefont {R.}~\bibnamefont {Takagi}},\ }\bibfield  {title} {\bibinfo {title} {One-shot manipulation of dynamical quantum resources},\ }\href {https://doi.org/10.1103/PhysRevLett.127.060402} {\bibfield  {journal} {\bibinfo  {journal} {Physical Review Letters}\ }\textbf {\bibinfo {volume} {127}},\ \bibinfo {pages} {060402} (\bibinfo {year} {2021})}\BibitemShut {NoStop}%
\bibitem [{\citenamefont {Yuan}(2019)}]{Yua19}%
  \BibitemOpen
  \bibfield  {author} {\bibinfo {author} {\bibfnamefont {X.}~\bibnamefont {Yuan}},\ }\bibfield  {title} {\bibinfo {title} {Hypothesis testing and entropies of quantum channels},\ }\href {https://doi.org/10.1103/PhysRevA.99.032317} {\bibfield  {journal} {\bibinfo  {journal} {Physical Review A}\ }\textbf {\bibinfo {volume} {99}},\ \bibinfo {pages} {032317} (\bibinfo {year} {2019})}\BibitemShut {NoStop}%
\end{thebibliography}%
\end{document}